\documentclass[letterpaper,11pt]{article}

\usepackage{jheppub}
\usepackage{mathtools,amssymb,braket,mathrsfs,tikz,subfig,xspace,xcolor,float,color,siunitx,comment,afterpage,multirow,array,hhline,amsmath,amsthm,framed,colortbl, bm}
\usepackage[countmax]{subfloat}

\newcommand{\be}{\begin{equation}}
\newcommand{\ee}{\end{equation}}

\renewcommand{\O}{\mathcal{O}}

\usepackage{tikz}
\usetikzlibrary{fit}
\tikzset{%
  highlight/.style={rectangle,rounded corners,fill=red!15,draw,
    fill opacity=0.5,thick,inner sep=0pt}
}

\usepackage{amsmath,amssymb,amsthm,cancel,hyperref,graphicx,xcolor}
\usepackage{mathrsfs,wasysym}
\usepackage{booktabs}
\usepackage{tikz}
\usetikzlibrary{calc}
\usepackage{placeins}

\usepackage{slashed}
\usepackage{graphicx}

\usepackage{soul}

\usepackage{array}

\usepackage{comment}

\usepackage{url}
\usepackage{xspace}
\usepackage{dsfont}
\usepackage{amssymb}
\usepackage{amsmath}
\usepackage{graphicx}
\usepackage{hhline}
\usepackage{physics}

\usepackage{color}
\definecolor{mit-red}{rgb}{0.64,.12,0.2}
\definecolor{darkred}{rgb}{1.0,0.1,0.1}
\definecolor{darkgreen}{rgb}{0.1,0.7,0.1}
\definecolor{darkblue}{rgb}{0.1,0.1,1.0}
\definecolor{pink}{rgb}{1.0, 0.0, 0.67}

\usepackage{amsmath}

\DeclareRobustCommand{\Sec}[1]{Sec.~\ref{sec:#1}}
\DeclareRobustCommand{\Secs}[2]{Secs.~\ref{sec:#1} and \ref{sec:#2}}
\DeclareRobustCommand{\App}[1]{App.~\ref{app:#1}}

\DeclareRobustCommand{\Tab}[1]{Table~\ref{tab:#1}}

\DeclareRobustCommand{\Fig}[1]{Fig.~\ref{fig:#1}}
\DeclareRobustCommand{\Figs}[2]{Figs.~\ref{fig:#1} and \ref{fig:#2}}

\DeclareRobustCommand{\Eq}[1]{Eq.~(\ref{eq:#1})}
\DeclareRobustCommand{\Eqs}[2]{Eqs.~(\ref{eq:#1}) and (\ref{eq:#2})}
\DeclareRobustCommand{\Reference}[1]{Ref.~\cite{#1}}

\newcommand{\Adam}{{\sc Adam}\xspace}

\begin{document}

\count\footins = 1000
\interfootnotelinepenalty=10000
\setlength{\footnotesep}{0.6\baselineskip}

\renewcommand{\arraystretch}{1.3}

\title{Resummed Distribution Functions: Making Perturbation Theory Positive and Normalized}

\author[a]{Rikab Gambhir,}
\author[b]{Radha Mastandrea}

\affiliation[a]{Department of Physics, The University of Cincinnati, Cincinnati, OH 45221, USA}
\affiliation[b]{The University of Chicago, Chicago, IL 60637, USA}

\emailAdd{gambhirb@ucmail.uc.edu}
\emailAdd{rmastand@uchicago.edu}

\abstract{
Fixed-order perturbative calculations for differential cross sections can suffer from non-physical artifacts: they can be non-positive, non-normalizable, and non-finite, none of which occur in experimental measurements.
We propose a framework, the \emph{Resummed Distribution Function} (RDF), that, given a perturbative calculation for an observable to some finite order in $\alpha_s$, will ``resum'' the expression in a way that is guaranteed to match the original expression order-by-order and be positive, normalized, and finite. 
Moreover, our ansatz parameterizes \emph{all} possible finite, positive, and normalized completions consistent with the original fixed-order expression, which can include N$^n$LL resummed expressions.  
The RDF also enables a more direct notion of perturbative uncertainties, as we can directly vary higher-order parameters and treat them as nuisance parameters. 
We demonstrate the power of the RDF ansatz by matching to thrust to $\mathcal{O}(\alpha_s^3)$ and extracting $\alpha_s$ with perturbative uncertainties by fitting the RDF to ALEPH data.
}

\maketitle

\section{Introduction}

 Unitarity is a powerful constraint on  predictions within a quantum field theory (QFT)~\cite{Cutkosky:1960sp, Eden:1966dnq}. 
At minimum, unitarity implies that the differential cross section $\dv{\sigma}{x}$ for a set of observables $x$ in a scattering process must be \textbf{normalized} and \textbf{non-negative}.
Normalization means that the integral $\sigma = \int_X dx\, \dv{\sigma}{x}$ is finite, or alternatively, the probability density function (PDF) $p(x) \equiv \frac{1}{\sigma}\dv{\sigma}{x}$ integrates to 1.
Non-negativity means that $p(x) \geq 0$ everywhere.
A cross section that fails to satisfy either property cannot be physical.

To predict cross sections from a QFT, we often use \emph{fixed-order perturbation theory}.
Given a perturbative parameter $\alpha \ll 1$ in our theory, we can write a \emph{fixed order} (FO) calculation at order $M$ as
\begin{align}
    p_{\rm FO}(x|\alpha) = p_0(x) + \alpha p_1(x) + \alpha^2 p_2(x) + ... + \alpha^M p_M(x) + \O\left(\alpha^{M+1}\right) ,
\label{eq:expansion}
\end{align}
where $p_m(x)$ are calculable through perturbative techniques (e.g. Feynman diagrams), and $M$ is the (finite) order of the calculation.
Ideally, at the series converges to $p_{\rm FO}$ as $M \to \infty$.
It may be that the perturbative series is not actually analytic (as is the case for many perturbative series in QFT) and instead is an asymptotic series.
We can then only hope that $p_{\rm FO}(x|\alpha) \approx \sum_m^M \alpha^m p_m(x)$ for some large enough $M$ or small enough $\alpha$.

However, at any finite order $M$, our physical perturbative predictions for $p(x)$ may violate unitarity by being non-normalizable or negative.
Moreover, the perturbative expansion can be spoiled if  $p_m(x) \sim \frac{1}{\alpha^m}$, which is especially the case in quantum chromodynamics (QCD) where large logarithms due to infrared and collinear divergences can arise.
Additionally, while the perturbative series will be an integrable distribution, it will rarely be a proper function, and it may contain objects such as $\delta$-functions or $+$-functions (which can technically restore unitarity or cancel divergences).
Experimental measurements of $p(x|\alpha)$ are typically true functions that are finite everywhere as nature carries out all-orders calculations, and no experimental histogram will contain these objects.

In this paper, we introduce a new ansatz, which we call the \emph{Resummed Distribution Function} (RDF), for parameterizing proper PDFs of random variables.
This ansatz is guaranteed to be positive, finite, and normalized, and thus satisfy unitarity. 
The RDF (which we denote as $q(x|\alpha)$ where $x$ can be either a single observable or a set of several observables) is an all-orders expansion in $\alpha$, and it effectively ``resums'' the fixed-order series using only unitarity and analyticity in $\alpha$ as consistency conditions. 
Given a fixed-order expansion for $p(x|\alpha)$ up to order $M$, our ansatz $q(x|\alpha)$ is guaranteed to match $p(x|\alpha)$ up to order $M$, meaning that it contains all of the information in the corresponding $p_{\rm FO}(x|\alpha)$. 
Through this ansatz, we effectively parameterize, in terms of free parameters, \emph{all} possible higher-order extensions of $p(x|\alpha)$ permissible by unitarity.
The RDF provides an automated approach for generating these potential ``all-orders resummations'' without any appeal to the structure of the theory, making it especially useful if a proper resummation is difficult or unavailable.

The RDF provides an efficient machine-learning-inspired\footnote{We say ``inspired'' because while the RDF is a universal probability estimator with parameters that can be numerically fit with gradient descent like many machine learning methods, our aim is to build a primarily analytical intuition for it.} framework to learn  normalized, finite, and conditional multivariable PDFs from analytic calculations, simulations, or real data, complementary to established machine-learning-and-adjacent methods for density estimation.
We emphasize the strong inductive bias of the RDF ansatz: while it is a universal PDF approximator, it is particularly useful for distributions that admit a perturbative expansion in $\alpha$, as is often the case in a QFT.
In contrast to normalizing flows\footnote{We strongly considered naming our ansatz the ``Re-normalizing flow''.} \cite{papamakarios2021normalizingflowsprobabilisticmodeling} and diffusion models \cite{ho2020denoisingdiffusionprobabilisticmodels}, our ansatz is extremely lightweight: it can fit to numeric samples in a matter of minutes, requiring only a small set ($\sim \mathcal{O}(10)$) of trainable parameters.
The RDF has a manifest expansion in $\alpha$ where terms of the expansion are directly parameterized, unlike normalizing flows and diffusion models where the expansion is not manifest or necessarily well-behaved.
Compared to non-parametric methods for PDF estimation such as Kernel Density Estimation \cite{kde}, our method can easily be made conditional on a parameter $\alpha$.

The RDF enables an alternative to the usual approach of estimating theoretical uncertainties using scale variations.
After matching the RDF to a calculation of a given order by fixing low-order parameters, we can treat the RDF's higher order parameters as nuisance parameters, in the vein of \Reference{Tackmann:2024kci}.
This, combined with the fact that the RDF is always a valid distribution and contains some all-orders information, makes it useful for fitting, e.g. $\alpha_s$ extractions from event shapes~\cite{Kluth:2006bw, Bethke:2006ac,ALEPH:2003obs, OPAL:2011aa, Abbate:2010xh, dEnterria:2022hzv, Benitez:2024nav}.
Our approach is all-orders and systematically flexible, so that with enough parameters it can eventually capture all variations.
Theoretical knowledge can be inserted by imposing a ``prior'' on the higher order RDF coefficients, and we argue that a prior is necessary since there is no strictly frequentist interpretation of perturbative uncertainties.
We demonstrate that very simple and reasonable priors can lead to reasonable perturbative uncertainties on $\alpha_s$ extractions using event shape calculations at $\O(\alpha_s^2)$ and $\O(\alpha_s^3)$ in LEP data. 

The rest of this paper is organized as follows. 
In \Sec{renormalizing_flow}, we define and construct the RDF as a unitary and analytic ansatz $q(x|\alpha)$, and we illustrate how the information from a fixed-order, pre-calculated PDF $p_\textrm{FO}(x|\alpha)$ expressed as a perturbative series in $\alpha$ can be encoded into the ansatz.
In \Sec{resummation}, we briefly discuss the all-orders aspects and interpretation of the RDF and the connection to logarithmic resummation in QCD.
In \Sec{toy_examples}, we apply the RDF to simple toy distributions to show off the analytic and numeric matching procedures developed in \Sec{renormalizing_flow} (this section can be skipped by readers primarily interested in QCD applications). 
In \Sec{qcd_observables}, we show RDF construction for QCD shape observables: single jet angularities, simultaneous jet angularities, and the event thrust.
Finally, in \Sec{nuisance_params}, we discuss how higher-order parameters of the RDF can be used as nuisance parameters to define perturbative uncertainties.
We perform semi-realistic fits to ALEPH~\cite{ALEPH:2003obs} thrust data using the RDF to extract $\alpha_s$ up to and including $\O(\alpha_s^3)$.
We conclude in \Sec{conclusions}. 
We provide supplemental plots in \App{supp_plots}, and details about numerics in \App{numerics}.

\section{Resummed Distribution Functions: A Normalized, Positive, and Finite Ansatz for PDFs}
\label{sec:renormalizing_flow}

Suppose we have random variables $t_1, t_2, ... t_k$ all living on the domain $[0, \infty)$.
We are interested in estimating the distribution $p(t_1, t_2, ..., t_k |\alpha)$ for some perturbative parameter $\alpha$.
If we have a random variable on a finite domain, $x \in [0,1]$, we can always convert it to $t \in [0, \infty)$ with the transformation:
\begin{align}
    t \equiv \log(\frac{1}{x}).    
\end{align}
We will always use $t$ to refer to variables on $[0, \infty)$ and $x$ to refer to variables on $[0, 1]$.\footnote{The logarithm in the conversion will be especially useful for QCD later, but this is still a general physics-independent ansatz.}

Then, the Resummed Distribution Function (RDF) $q(t_1, t_2, ..., t_k|\alpha)$ is given by:
\begin{align}
    q(t_1, t_2, ..., t_k | \alpha) &= f_1(t_1, \alpha) \times ... \times f_k(t_1, t_2, ..., t_k, \alpha) \nonumber\\
    &\times \exp\left[-\int_0^{t_1}dt_1' \, f_1(t'_1, \alpha) \right]\times ... \times \exp\left[-\int_0^{t_k}dt_k' \, f_k(t_1, t_2,...,t_k', \alpha) \right], \label{eq:renormalizing_flow_multi}
\end{align}
where the functions $f$ are further given by functions $g$:
\begin{align}
    f_j(t_1, t_2, ... , t_j, \alpha) = \exp[-g_j(t_1, t_2, ... , t_j, \alpha)].
    \label{eq:f}
\end{align}
Here, the functions $g$ can be \emph{any} functions that (a) are bounded from above as any $t_i \to \infty$ and (b) are either analytic in $\alpha$ or have single-log dependence on $\alpha$.
This parameterization ensures that the $f$ functions are always positive and analytic in $\alpha$, which is necessary for our ansatz to be a valid and analytic PDF (which we discuss further in the following section).\footnote{Note that the functions $f$ and $g$ are ordinary functions of $t$ and $\alpha$, not PDFs of $t$ conditioned on $\alpha$.}
In \Eq{renormalizing_flow_multi}, the $j$'th integral is only integrating over the $t_j$ argument of $f_j$, and all other arguments are not integrated over.
The ansatz is completely specified by a choice of $g$ functions.
Any choice of $g$ functions satisfying the above two conditions is a valid PDF, and all PDFs correspond to some choice of $g$.

The univariate case of the RDF will be of special interest.
In this case, we have
\begin{align}
    q(t|\alpha) &= f(t, \alpha) \cdot \exp\left[-\int_0^t dt'\, f(t', \alpha)\right] \nonumber\\
    f(t, \alpha) &= \exp[-g(t, \alpha)],\label{eq:renormalizing_flow}
\end{align}
where, like above, $g$ is any real function that is bounded from above and is analytic up to single logarithms in $\alpha$.

In the rest of this section, we will first justify the form of this ansatz. 
Then, we will show how $g$ can be selected to preserve perturbative information from either an analytic or a numeric calculation.

\subsection{Univariate Taylor-Expandable Probability Distributions}

We first demonstrate why the univariate RDF of \Eq{renormalizing_flow} parameterizes all univariate probability densities.
Consider the cumulative distribution function (CDF), obtained by integrating $q(t|\alpha)$ from $0$ to $t$:
\begin{align}
    Q(t | \alpha) &= 1 - e^{-\int_0^t dt'\, f(t', \alpha)}. \label{eq:Q}
\end{align}
Since $g$ is bounded from above, $f = e^{-g}$ is strictly positive and has a diverging integral in $t$.
Therefore, $Q(t|\alpha)$ is monotonic and $Q(\infty|\alpha) = 1$, which implies that $Q$ is a valid and normalized CDF.  
Given any probability density $q(t|\alpha)$ with corresponding CDF $Q(t|\alpha)$, we may always write: 
\begin{align}
    f(t,\alpha) &= -\partial_{t}\log(1 - Q(t| \alpha)) \nonumber\\
    &= \frac{q(t|\alpha)}{1 - Q(t|\alpha)}, \label{eq:matching1}
\end{align}
demonstrating that every positive function $f$ with a diverging integral maps one-to-one with a valid probability distribution $q$.

If $q$ is analytic in $\alpha$, then $f$ is too by \Eq{matching1} --- this analyticity only holds if $Q(t |\alpha) < 1$, but this is automatically true almost everywhere in $t$ if $Q$ is a valid CDF.
Note that $Q$ is analytic if $q$ is analytic, since it is equal to the integral of $q$ over $t$ and is thus decoupled from analyticity in $\alpha$.
\Eq{matching1} is straightforward to Taylor expand in $\alpha$, and thus we can algorithmically convert the Taylor coefficients of $q$ into the Taylor coefficients of $f$, and therefore the coefficients of $g = -\log(f)$.
We will see in \Sec{analytic_matching} how to do this explicitly.

As stated previously, $g$ can have either analytic or single-log dependence on $\alpha$.
We may write:
\begin{align}
    g(t, \alpha) = - \log(g^*(t, \alpha)) + g_{\rm Analytic}(t, \alpha),
    \label{eq:base_rdf}
\end{align} 
for functions $g^*$ and $g_{\rm Analytic}$.
The minus sign is a convention motivated by the matching procedure in \Sec{analytic_matching}.
Here, $g_{\rm Analytic}(t, \alpha)$ is analytic in $\alpha$.
We must have $g^*$ be analytic in $\alpha$ \emph{and} positive over the full domain, such that $f = \exp(-g)$ is positive. 
While this additional constraint on $g^*$ is somewhat inconvenient, it is necessary to have a single-log dependence on $\alpha$ such that we can have PDFs that go to zero as $t \rightarrow 0$ and $\alpha \rightarrow 0$.

Thus we write the full univariate RDF as 
\begin{align}
q(t|\alpha) = g^*(t,\alpha) \exp\left[-g_\textrm{Analytic}(t, \alpha) - \int_0^t dt'\, g^*(t',\alpha) \exp\left(-g_\textrm{Analytic}(t', \alpha) \right) \right].
\end{align}
We can also write the corresponding CDF as:
\begin{align}
    Q(t | \alpha) = 1 - \exp\left[ - \int_0^t dt'\, g^*(t',\alpha) \exp\left(-g_\textrm{Analytic}(t', \alpha) \right) \right] \label{eq:q_CDF}.
\end{align}
Working with either the CDF or PDF is equivalent within the RDF framework, and we will find it convenient to switch between the two. 

Up to a given order, we may write $g_{\rm Analytic}$ as a polynomial in $\alpha$, where the coefficients are arbitrary functions of $t$. 
We may do similarly for $g^*$, as long as it is greater than zero.
While $g$ can be \emph{any} function satisfying the above properties (such as a learnable neural network), we will find it convenient for the purposes of this paper to parameterize each part of $g$ as a polynomial in $t$:
\begin{align}
    g^*(t, \alpha),\,\,\, g_{\rm Analytic}(t, \alpha) = \sum_{m=0,n=0}^{M,N} g_{mn}\alpha^m t^n \Theta(t - \theta_{mn}), \label{eq:g_ansatz}
\end{align}
where $g_{mn}$ are coefficients, and $N$ is the highest power of $t$ considered. 
The only requirement is that $g_{mN} < 0$ for normalization so that $g$ remains bounded from above.
With this parameterization, each part $g$ is a series of polynomials in $t$ and $\alpha$, modulated by $\Theta$-functions in cases where $t$ does not necessarily start at zero.\footnote{We draw a comparison here to \Reference{Assi:2025ibi}, which also involves determining parameters in an exponential to match to a calculation. However, \Reference{Assi:2025ibi} matches to already-resummed moments, and our $g$'s are inside of an exponential-integral-exponential structure rather than a plain single exponential.}

\subsection{Multivariate Taylor-Expandable Probability Distributions}
We now justify the full multivariate ansatz of \Eq{renormalizing_flow_multi} by building off of the univariate case.
A multivariate distribution $p(t_1, t_2, ..., t_k |\alpha)$ can always be decomposed as a chain of conditional distributions autoregressively:
\begin{align}
    p(t_1, t_2, ..., t_k |\alpha) = p(t_1 | \alpha) \cdot p(t_2|t_1, \alpha) \cdot ... \cdot p(t_k|t_1,t_2,...,t_{k-1}, \alpha).
\end{align}
The ordering of the $t_i$ is arbitrary, and thus this decomposition is \emph{not} unique.\footnote{Alternatively, one can choose to take an appropriately normalized sum over all possible orderings, but this is expensive computationally and does not offer any great advantage in our case.}

This decomposition allows us to apply \Eq{renormalizing_flow} separately to each individual term. 
The first term is identical to \Eq{renormalizing_flow}.
Each subsequent term is slightly nontrivial, because it involves additional conditional parameters.
However, this is not an issue:
in \Eq{renormalizing_flow_multi}, we only perform the integral over the random variable of interest, and not the conditional parameters.
For each $p(t_k | t_1 ... t_{k-1}, \alpha)$, we can construct an $f(t_1, ... t_k, \alpha)$, exactly as we   constructed $f(t, \alpha)$ from $p(t |\alpha)$ in the univariate case. 
When the $f$'s are integrated over in the RDF ansatz for each individual term, only the random variable $t_k$ is integrated over; all others are conditional.
With this, we have constructed the full RDF ansatz of \Eq{renormalizing_flow_multi}.

\subsection{Analytic Matching}\label{sec:analytic_matching}

Given a parameterization for the ansatz $q(t|\alpha)$, we would like to \emph{match} it order-by-order to a preexisting, theoretically-derived, fixed-order PDF $p_{\rm{FO}}(t|\alpha) = \sum_m^{M} p_m(t) \alpha^m$, defined up to a given order $M$.
Through this matching procedure, we will encode all information, up to that order $M$, that is contained in $p_{\rm{FO}}(t|\alpha)$ into the ansatz by carefully choosing the parameterized function $g$ such that: 
\begin{align}
q(t|\alpha) = p_{\rm FO}(t | \alpha)  + \mathcal{O}(\alpha^{M+1}).
\label{eq:ansatz_vs_FO}
\end{align}
The higher-order information in the ansatz will enforce normalization, positivity, and finiteness, effectively performing an ``all-orders resummation''. 
For the following analytic matching procedure, we assume that we have access to an \emph{explicit} expression for $p_{\rm FO}(t | \alpha)$ and the coefficient functions $p_m$ as defined in \Eq{expansion}.
Moreover, we require that the $p_m$ are indeed the Taylor coefficient functions of a well-defined distribution; that is, there exists an actual all-orders distribution the $p_m$ eventually converge to.\footnote{This is \emph{not} the case if, for example, one has $p_0(t) = 1$ and $p_{m}(t) = 0$ for all $m > 0$. This does not converge to a valid distribution over $t$.}
Note that, in actual QFT calculations and especially QCD, this assumption may not be true: not only is there nonperturbative physics, but often perturbative expansions are only asymptotic series that do not actually converge. 
These issues are not unique to our method and we do not claim to solve either problem --- the RDF is only as good as the perturbation theory used to derive $p_{\rm FO}$.

We start by taking the log of \Eq{matching1}:
\begin{align}
    \log(f(t, \alpha)) = \log(q(t|\alpha)) - \log(1 - Q(t| \alpha)).
\label{eq:tmp}
\end{align}
By definition, $g(t,\alpha) = -\log(f(t,\alpha))$, so we can substitute that into the left hand side of \Eq{tmp}.
Substituting \Eq{ansatz_vs_FO} into the right hand side of \Eq{tmp}, we recover
\begin{align}
    g(t, \alpha) = -\log(p_\mathrm{FO}(t|\alpha) + \mathcal{O}(\alpha^{M+1})) + \log(1 - P_\mathrm{FO}(t| \alpha) + \mathcal{O}(\alpha^{M+1})).
\label{eq:g_matching}
\end{align}
where $P_\mathrm{FO}$ is the CDF corresponding to $p_{\rm{FO}}$.
Here, we keep careful track of the $\mathcal{O}(\alpha^{M+1})$ terms inside the logarithms.

To extract the $g$ functions, we start by dividing out the lowest power of $\alpha$ from the fixed-order PDF, i.e.
\begin{align}
p_{\rm FO}(t |\alpha) = p_{m^*}(t)\alpha^{m^*}\left(1 + \sum_{m > m^*}^M \frac{p_m(t)}{p_{m^*}(t)}\alpha^{m-m^*} + \mathcal{O}(\alpha^{M + 1 - m^*})\right).
\label{eq:divide_out}
\end{align}
Notice the $\mathcal{O}(\alpha^{M + 1 - m^*})$ inside the parentheses --- since we have divided out $\alpha^{m^*}$, the expression inside the parentheses is an $\mathcal{O}(\alpha^{M - m^*})$ expression. 
We highlight this as it will be essential in our $\alpha$-counting.
Substituting \Eq{divide_out} into \Eq{g_matching}, we have
\begin{align}
    g(t, \alpha) &= -\log(p_{m*}(t)\alpha^{m^*})  \nonumber\\ &\quad- \log(1 + \sum_{m > m^*}^M \frac{p_m(t)}{p_{m^*}(t)}\alpha^{m-m^*}+ \mathcal{O}(\alpha^{M + 1 - m^*})) \nonumber\\ &\quad+ \log(1 - \int_0^t  dt'\sum_{m=m^*}^M p_m(t')\alpha^m + \O(\alpha^{M+1})). \label{eq:derivation_logs} 
\end{align}
We see that the leading term $p_{m^*}(t)\alpha^{m^*}$ maps directly to the single-log $g_*$ term in the ansatz from \Eq{base_rdf}.

We now wish to Taylor expand terms 2 and 3 to continue getting the ansatz into the form of \Eq{base_rdf}. 
We cannot Taylor expand the first term $-\log(p_{m*}(t)\alpha^{m^*})$, which acts as a common prefactor and encodes the leading small-$\alpha$ behavior.
Since the entire ansatz is meant to capture the correct scaling behavior up to $\alpha^M$, this means that any contributions of orders \textit{higher} than $\alpha^{M-m^*}$ in terms 2 and 3 will not contribute when combined with term 1. 
This does not affect term 2.
However, term 3 changes: we adjust the $\mathcal{O}(\alpha^{M + 1 })$ to $\mathcal{O}(\alpha^{M + 1 - m^*})$, and we adjust the bounds of the sum to only go up to $M - m^*$, rather than $M$.

Taylor-expanding the logarithms and collecting terms 2 and 3, we have
\begin{align}
   g(t,\alpha)  &= -\log(p_{m*}(t)\alpha^{m^*}) \nonumber\\
    &\quad+ \left[ \sum_{k \geq 1}\frac{1}{k}\left(\left[-\sum_{m > m^*}^M \frac{p_m(t)}{p_{m^*}(t)}\alpha^{m-m^*} \right]^k - \left[ \sum_{m=m^*}^{M-m^*} \int_0^t dt'\, p_m(t') \alpha^m \right]^k\right) + \O\left(\alpha^{M+1-m^*}\right)\right].
\label{eq:full_matching}
\end{align}
In expanding the logarithms, we make use of the fact that $\alpha \frac{p_m(t)}{p_{m^*}(t)}$ are small for each $m$.
This is, in effect, our ``resummation''. 
Rather than expanding around just $\alpha$, we instead expand around these effective parameters.
In the case where the $p_m(t)$ are logarithms, this is superficially similar to (but not necessarily the same as) standard logarithmic resummation techniques in QCD, as discussed further in \Sec{resummation}.
Importantly, the $\O(\alpha^{M+1-m^*})$ is inside the square brackets, not outside, because only the bracketed term is analytic.

Comparing \Eq{full_matching} with \Eq{g_matching}, we see that $g^*(t, \alpha)$ is exactly $p_{m^*}(t)\alpha^{m^*}$, and that $g_{\rm Analytic}(t, \alpha)$ is given by the second line of \Eq{full_matching}. 
Note that the second line of \Eq{full_matching} is explicitly a power series in $\alpha$. Thus if we express $g_{\rm{Analytic}}(t, \alpha)$ as a power series in $\alpha$, e.g. $\sum_M g_m(t)\alpha^m$, the $g_m(t)$ can be automatically extracted by matching powers of $\alpha$.  
Importantly, the analytic part of \Eq{full_matching} needs to only be computed up to and including $\mathcal{O}(\alpha^{M-m^*})$, \emph{not} $\mathcal{O}(\alpha^{M})$.
This means any contributions from the infinite power series in $k$ can be ignored beyond this point.
We will choose to set all $\mathcal{O}(\alpha^{M+1-m^*})$ terms to zero by default when we perform matching --- we emphasize, however, that this is a choice, since in principle anything is allowed without spoiling the matching.
While only up to $g_M(t)$ is required to match, in general $g_m(t)$ for $M \geq m$ can be be nonzero --- though care should be taken, as the radius of convergence is finite and additional terms may cause $q(t|\alpha)$ to diverge from $p(t|\alpha)$ even if they formally match.

\subsection{Numeric matching}
\label{sec:numeric_matching}
In the previous section, we showed how to encode the information from a fixed-order PDF into the RDF functional form. 
However, we can still construct an RDF solution for a given observable even if we do not have the fixed-order PDF.
All we need is a histogram of the observable, either from experimental data\footnote{If one has access to data, then it is better to match to the RDF directly rather than doing a fixed-order matching, as we will do in \Sec{nuisance_params}.} or a fixed-order Monte Carlo (MC) computational tool such as \textsc{MadGraph}~\cite{Alwall:2011uj} or \texttt{EERAD3}~\cite{Aveleira:2025svg} that is expected to be valid to some order in $\alpha$ in a known region of the domain of $t$.
We can then carry out a numeric matching procedure to find a solution for $g(t,\alpha)$ that best fits the data or MC across this $t$ domain.

In this work, we will assume that both $g^*(t,\alpha)$ and $g_{\rm{Analytic}}(t,\alpha)$ can be expressed as polynomials in $t$ (recall that $g(t, \alpha) = g_{\rm Analytic}(t, \alpha) - \log(g^*(t, \alpha))$), as in \Eq{g_ansatz}, though with minor modifications for numeric stability described below.
This choice allows us to numerically learn a suitable ansatz through a small set of coefficients $g^*_{mn}$ and ${g_\mathrm{Analytic}}_{mn}$.


The construction of the numeric RDF is as follows.
We initialize two matrices of coefficients $g^*_{mn}$ and ${g_\mathrm{Analytic}}_{mn}$, each of size $M\times N$.
$M$ corresponds to the order in $\alpha$ which we are matching to, and $N$ corresponds to the desired maximum power in $t$.
We may optionally learn two vectors of $\theta$-function components $\theta^*_m$ and ${\theta_\mathrm{Analytic}}_m$, which allow us to parameterize solutions that do not go to zero at $t = 0$.
As in the analytic case, we specify a minimum power $m^*$ in $\alpha$ for the RDF.
The functional form to numerically fit is then given by
\begin{align}
g(t, \alpha) = &-\log(\sum_{m=m^*}^{M}\frac{\alpha^{m}}{m!}\left|\sum_{n=0}^{N}g^*_{mn}\frac{t^n}{n!} \Theta_{T_m}(t-\theta^*_{m})\right|_{T_m}) \nonumber\\
&+ \sum_{m=0,n=0}^{M-m^*,N} {g_\mathrm{Analytic}}_{mn}\frac{t^n}{n!}\frac{\alpha^{m}}{m!} \Theta_{T_m}(t-{\theta_\mathrm{Analytic}}_{m}),\label{eq:numeric_ansatz}
\end{align}
where the first term corresponds to $\log(g^*(t, \alpha))$ and the second term to $g_{\rm Analytic}(t, \alpha)$.  
The numeric RDF differs from the analytic formula of \Eq{g_ansatz} in a few minor ways.
First, we allow the argument of the logarithm to be an arbitrary polynomial in $\alpha$, rather than using only the lowest power.
We also force it to be strictly positive using with an absolute value function.\footnote{Why absolute value and not anything else, e.g. a smoothed-out ReLU~\cite{agarap2019deeplearningusingrectified} to enforce positivity? We tried, and it was not as numerically stable.}
Second, we include explicit factorial scaling of the $g_{mn}$ coefficients for numerical stability.
Lastly, we ``smooth out'' the $\Theta$-functions and absolute value functions, as indicated by the subscript $T$.
We do this with the replacements $\Theta_T(x) \to \sigma(x; T) = \frac{1}{1 + e^{-x/T}}$ and $|x|_T \to 2\sigma(\frac{x}{T})x -x$, with learnable parameters $T^*_m$, $T_{{\rm Analytic}, m}$, and $T_{{\rm Abs}, m}$ respectively per each $m$.
This ensures that the derivatives of the RDF are well-defined at all points in $t$-space, which is especially important for the $\theta_m$'s to be learnable via gradient descent.
All in all, we fit seven different objects: $g^*_{mn}$, ${g_\mathrm{Analytic}}_{mn}$, $\theta^*_m$, ${\theta_\mathrm{Analytic}}_m$, $T^*_m$, $T_{\mathrm{Analytic}, m}$, and $T_{\textrm{Abs}, m}$.
For convenience, we will refer to this set of seven learnable objects as $\phi$.

Importantly, if $m^* > 0$, then the $m$th row of $g_{\rm Analytic}$ contributes to $f$ at $\O(\alpha_s^{m^*})$ higher than the corresponding $m$'th row of $g^*$.
Therefore, if we are working only to some fixed order $M$, we are free to completely ignore the $M-m^*$'th through $M$'th rows of $g_{\rm Analytic}$, as these will not contribute anyways except at higher-orders.
We will choose to freeze these rows of $g_{\rm Analytic}$ at zero, and (by choice) we set the corresponding rows of $\theta_{\rm Analytic}$ and $T_{\rm Analytic}$ to -1.0 and 0.1 respectively.

The actual matching procedure is implemented in \texttt{JAX} \cite{jax2018github}.
For every training epoch, we sample some number of $\alpha$'s (corresponding to one batch) from some underlying distribution.
For each $\alpha$, we pull the corresponding PDF that we want to fit to (i.e. from MC simulation) across a specified fitting domain [$t_\mathrm{min}$, $t_\mathrm{max}$].
We then calculate the Taylor expansion of the RDF around $\alpha $ up to a pre-specified matching order (ideally the same order as the fixed-order PDF). 
We then calculate the MSE (mean-squared error) loss between the (binned) MC PDF and the Taylor-expanded RDF as
\begin{align}
\begin{split}
\mathrm{Loss}(\alpha, \phi) = \frac{1}{2}\sum_{\textrm{Bin}_i}\frac{| \sum_{m=m*}^M\frac{1}{m!}\frac{\partial^m\mathrm{RDF}}{\partial\alpha^m}(\textrm{Bin}_i, \alpha, \phi)- \textrm{Target}(\textrm{Bin}_i)|^2}{\textrm{Error}(\textrm{Bin}_i)^2}. 
\label{eq:numeric_mse_loss}
\end{split}
\end{align}
The binwise error in the denominator comes from uncertainties on the data or MC.
We then backpropagate to calculate gradients of the loss with respect to the parameters and update the $g_{mn}$ and $\theta_m$ arrays with a gradient optimizer, in our case \Adam~\cite{adam}.

We state the exact values for all hyperparameters (e.g. batch size, number of epochs, learning rate) used to generate the numeric RDF for every observable we show in this work in \App{training_hyperparams}.
However, we stress that hyperparameters were not heavily optimized, and they are quite similar across all observables.
To initialize the parameters, we use a random ``reroll'' procedure, the details of which are available in \App{initialization}.

\section{What, When, and Why RDFs?}\label{sec:resummation}

In this section, we hope to make clear what the RDF can and cannot do. 
It is important to be careful about the interpretation of the RDF, as it is \emph{not} a method for obtaining extra information beyond unitarity for ``free'' or for performing automatic logarithmic resummation.
At the lowest level, the \emph{only} thing an RDF does is answer the question: ``\textit{Given a fixed-order calculation, plus the additional knowledge that the all-orders distribution should be a proper probability distribution, what are all probability distributions consistent with that calculation?}''
The non-matched part of $g$ is infinitely flexible, and so without any additional information, further calculation, or a Bayesian prior, one cannot pick out \emph{which} of these distributions is the correct one.

First, a disclaimer: the RDF is manifestly analytic (and therefore convergent) in the parameter $\alpha$, but generic perturbative expressions in QFT are expected to be asymptotic expansions and diverge~\cite{Dyson:1952tj}.
The RDF cannot capture this feature; rather, the RDF can only answer what is perturbatively consistent with the given fixed-order expression.
In other words, ``\textit{Given that we believe in perturbation theory, what are the possible all-orders distributions?}''
One cannot, for example, use the RDF to reproduce $e^{-\frac{1}{\alpha_s}}$-type singularities that one would expect due to resurgence.
Similarly, the RDF as specified has no knowledge whatsoever of nonperturbative physics --- the method is purely perturbative.

However, there is some information gain due to the unitarity assumption: unitarity is an all-orders statement, so the RDF effectively provides a constraint on what the higher-order terms could be. 
This is indeed a nontrivial constraint: for example, after matching with an RDF to order $M$, and Taylor-expanding to see how the higher order $q_m$ for $m > M$ depend on the choice of $g$, one will \emph{never} find that $q_m = 0$ for all $m > M$\footnote{Except in the trivial case where the original fixed-order expression happened to have no $\alpha$-dependence and was normalized to begin with.}, regardless of the choice of $g$, since that would violate unitarity. 
No matter what the original $p_m$ are, the higher-order coefficients must always conspire to make the entire distribution positive and normalized, which a generic all-orders guess may not do.

The RDF is an expansion in $\alpha^{m-m^*} \frac{p_m(t)}{p_{m^*}(t)}$, where $p_m(t)$ is the order-$m$ term in the $\alpha$ expansion and $p_{m^*}$ is the first nontrivial term in the expansion.
Specifically, we assume in \Eq{full_matching} that $\alpha^{m-m^*} \frac{p_m(t)}{p_{m^*}(t)} \ll 1$ in expanding the logarithms.
This is the ``resummation'' in ``Resummed Distribution Function'': it is a reorganization of the perturbative series in $\alpha^{m-m^*} \frac{p_m(t)}{p_{m^*}(t)} \ll 1$ for each $m$.
If the $p_m$ contain large logarithms, the RDF representation encodes them to all orders through the $g$-functions, but \emph{only} those particular logarithmic structures implied by the fixed-order input.
This should not be confused with the canonical N$^n$LL logarithmic resummation structure in QCD derived from factorization and renormalization group (RG) methods~\cite{Bauer:2000yr, Banfi:2004yd}.
The choice of the higher-order components of $g$ effectively amounts to guessing this structure.
It may be possible to extend the RDF to include this information (by e.g. imposing explicit $\mu$-dependence or factorization structure) in potential future work, but we do not pursue this here and content ourselves with fixed-scales and no assumed internal structure.
This is resummation in the broad sense of formally reorganizing a perturbative series, of which the usual N$^n$LL resummation is one specific type.

However, it is still highly tempting to draw a connection between the RDF and logarithmic resummation in QCD, especially since our ansatz consists of exponentiated polynomials in $\alpha$ and $\log(1/x)$.
Such a connection must be made with care and a few caveats.
A logarithmic resummation for the CDF $P$ of an observable $x$ takes the form~\cite{Catani:1992ua, Banfi:2004yd}\footnote{We have altered the notation and sign conventions of \Reference{Catani:1992ua} to better match our own. What they call ``$G_{nm}$'', we call ``$-F_{mn}$'' (note the swap between $m$ and $n$), since these coefficients are best associated with the integrals of our $f$-functions, $F = \int dt\, f$.}:
\begin{align}
    P(x|\alpha) = (1 + \sum_{m = 1}^\infty C_m \alpha^m) \times \exp[-\sum_{m=1}^\infty\sum_{n = 1}^{m+1} F_{mn} \alpha^m \log^{n}(1/x)] + R(x), \label{eq:logarithmic_resummation}
\end{align}
where $C_m$ and $F_{mn}$ are coefficients and $R$ is a remainder function that goes to $0$ as $x \to 0$.
A \emph{leading log} (LL) calculation is one that includes \emph{all} terms of the form $\alpha^m \log^{m+1}(1/x)$, a \emph{next-to-leading log} (NLL) calculation is one that includes \emph{all} terms of the form $\alpha^m \log^{m}(1/x)$, and so on to define N$^n$LL. 
A related concept is the \emph{double log} (DL) calculation, which includes the $\alpha \log^{2}(1/x)$ piece in the exponent at fixed coupling (unlike a true LL calculation which includes an entire tower of terms and the effects of running couplings).
We can compare the formula for the logarithmic resummation of the CDF to the integral of the RDF, which is of the form (converting back to $x$-coordinates rather than $t$)
\begin{align}
    Q(x |\alpha) = 1-\exp[-F(\log(1/x),\alpha)],\,\,\, \text{where}\,\, F(t, \alpha) = \int_0^t dt' f(t, \alpha) .
\end{align}
Assuming that the $g$ functions are polynomial in $\alpha$ and $t$ as in \Eq{g_ansatz}, then $F$ can be approximated by a polynomial in $\alpha$ and $t$ with coefficients $F_{mn}$.
If we ignore all non-singular terms in the original fixed-order expansion before matching (i.e., if we are working in the soft-collinear limit of QCD), then not only will there be no remainder term as there is in \Eq{logarithmic_resummation}, but $Q$ will also be comprised solely of logarithmic contributions in $x$.
In these cases, if the correct $F_{mn}$ coefficients are known, the RDF genuinely captures N$^n$LL effects.
That is, an N$^n$LL calculation ``fits'' within the RDF framework as a special case.

However, \emph{extreme} care must be taken in interpreting an RDF matching as an N$^n$LL calculation. 
First, one must guarantee that \emph{all} logarithms of the desired order are present in the fixed-order calculation --- for example, at NLL, this includes effects due to the running of $\alpha_s$ and non-global logarithms, which are nontrivial to treat. 
Without this, there is no guarantee that matching to a fixed-order calculation will determine all the $F_{mn}$ needed to claim N$^n$LL accuracy.
Often, one only has the ability to compute a subset of the logarithms at any given order (see e.g. ``Modified-Leading Logarithm (MLL) calculations''~\cite{Dokshitzer:1992jv} that miss some logarithms, or non-global logs~\cite{Dasgupta:2001sh}), and we will see in \Sec{multi_angularity} an explicit example where ``leading order'' does not imply ``leading logs'' in the case of simultaneous observables. 
Second, the non-matched part of $g$ controls all higher terms in the expansion.
With no additional information or prior, this is effectively a random choice on the coefficients of the higher-order terms, which represents only a partial summation of all logarithms of that order.
One must choose $g$ carefully to set these terms to zero if the goal is to compare to an N$^n$LL calculation and look solely at the logarithms up to a given order.

If we \emph{already} have a fully logarithmically resummed calculation for an observable that we trust and that is already a finite, normalized, positive distribution, or if we do not care about resummation or unitarity, is there still any value to the RDF framework? 
We believe that the answer is yes.
As we will explore further in \Sec{nuisance_params}, the higher-order parameters of the RDF can be used to parameterize arbitrary perturbative uncertainties, analogous to scale variations.
Even just a random variation of higher-order parameters of the $g$-function can give a qualitative sense of the perturbative convergence of the calculation without any reference to arbitrary scales, as we will make use of in \Secs{toy_examples}{qcd_observables} extensively.

\section{Matching to Toy Examples}
\label{sec:toy_examples}

In this section, we demonstrate the RDF on toy models with known analytic forms as a ``warm up'' to show the method in action.
The purpose of this section is to (a) demonstrate that the RDF works as advertised and (b) to give a taste of how to use it.
This section may be skipped by readers primarily interested in physics applications.

We will explore two toy models: the exponential distribution and the Rayleigh distribution:
\begin{align}
    p_{\rm Exponential}(t | \alpha) &= \alpha e^{-\alpha t} \nonumber\\
    p_{\rm Rayleigh}(t | \alpha) &= \alpha t e^{-\alpha \frac{t^2}{2}}.
\label{eq:toys_closed_form}
\end{align}
The Rayleigh distribution is chosen in part due to its resemblance to the Sudakov factor for $t = \log(\frac{1}{x})$.
In $x$-space, these observables are:
\begin{align}
        p_{\rm Exponential}(x | \alpha) &= \alpha x^{\alpha - 1} \nonumber \\
    p_{\rm Rayleigh}(x | \alpha) &= \alpha \frac{\log(\frac{1}{x})}{x}  e^{-\alpha \frac{\log^2(\frac{1}{x})}{2}}.
\end{align}
For the purpose of these toy models, we will assume that we \emph{do not} know the true forms of these distributions. 
Rather, we will only assume we know up to a finite order in $M$:
\begin{align}
    p_{\rm FO}^{(\rm Exp)}(t | \alpha) &= \alpha \sum_{m = 0}^{M-1} \frac{1}{m!}(-\alpha t)^m  + \O(\alpha^{M+1})\nonumber\\
    p_{\rm FO}^{(\rm Ray)}(t | \alpha) &= \alpha t \sum_{m = 0}^{M-1} \frac{1}{m!}\left(-\alpha \frac{t^2}{2}\right)^m   + \O(\alpha^{M+1}).
\label{eq:toy_FO_pdfs}
\end{align}
These fixed-order distributions are \emph{not} positive, finite, or normalizable for any nonzero value of $\alpha$.
We will see in the following studies how the RDF can cure these pathologies both analytically and numerically, without having to make use of higher-order knowledge.

\subsection{Analytic matching}\label{sec:toys_analytic}

We first show how the analytic matching procedure outlined in \Sec{analytic_matching} can be used to extract the functions $g^*(t, \alpha)$ and $g_\textrm{Analytic}(t, \alpha)$ on the two toy examples.

We begin with the exponential example. We take the fixed-order exponential expression (the top line of \Eq{toy_FO_pdfs}) for some finite $M$ and compare it to \Eq{divide_out}, which defines the start of the matching procedure.
A direct comparison of terms tells us that $p_{m^*}(t) = 1$, $m^* = 1$, and $p_m(t) = \frac{(-t)^{m-1}}{(m-1)!}$. 
We may then extract the $g$ functions by implementing \Eq{full_matching}:
\begin{align}
    g(t, \alpha) &= -\log(\alpha) + \left\{ \sum_{k}\frac{1}{k}\left(\left[-\sum_{m > 1}^{M-1}  \frac{(-\alpha t)^{m-1}}{(m-1)!} \right]^k - \left[ \sum_{m=1}^{M-2} \int_0^t dt'\,  \frac{(-t')^{m-1}}{(m-1)!} \alpha^m \right]^k\right) + \O\left(\alpha^{M}\right)\right\} \\
     &\downarrow \text{(Carrying out the integral in the third term)}\nonumber\\
    &= -\log(\alpha) + \left\{ \sum_{k}\frac{1}{k}\left(\left[-\sum_{m = 2}^{M-1}  \frac{(-\alpha t)^{m-1}}{(m-1)!} \right]^k - \left[ \sum_{m=1}^{M-2} -\frac{(-\alpha t)^{m}}{m!} \right]^k\right) + \O\left(\alpha^{M}\right)\right\} \\
       &= -\log(\alpha) + \left\{ \sum_{k}\frac{(-1)^k}{k}\left(\left[\sum_{m' = 1}^{M-2}  \frac{(-\alpha t)^{m'}}{(m')!} \right]^k - \left[ \sum_{m=1}^{M-2} \frac{(-\alpha t)^{m}}{m!} \right]^k\right) + \O\left(\alpha^{M}\right)\right\}
       \\
       &= -\log(\alpha) + \O\left(\alpha^{M}\right).
    \label{eq:exponential_matching}
\end{align}
The entire sum is $\O(\alpha^{M})$ and can therefore (in our matching convention) simply be set to zero, since we only have to match the analytic part of $g$ to order $M+1-m^* = M$.
We have extracted 
\begin{align}
\begin{split}
&g^*(t,\alpha) = \alpha\\
&g_\textrm{Analytic}(t, \alpha) =0+ \O(\alpha^M).\label{eq:exponential_g}
\end{split}
\end{align} 
Thus we can  completely specify the full RDF as
\begin{align}
\begin{split}
    q^{(\textrm{Exp})(M)}(t | \alpha) &= g^*(t,\alpha) \exp\left[-g_\textrm{Analytic}(t, \alpha) - \int_0^t dt'\, g^*(t',\alpha) \exp\left(-g_\textrm{Analytic}(t', \alpha) \right)  \right]\\
     &= \alpha \exp\left( - \alpha t  \right) + \mathcal{O}(\alpha^{M+1}). \label{eq:exponential_completion}
\end{split}
\end{align}
For this example, the matching procedure seems to have done better than matching the exponential expression up to order $M$ --- it has actually given us the all-orders PDF!
However, we are free to choose any $\O(\alpha^M)$ expression for $g_\textrm{Analytic}(t, \alpha)$, and it is only due to the simplicity of the example that choosing these higher-order terms to be zero happens to reproduce the true distribution.
(In \Fig{exponential_completion} below, we will show what happens when this choice is not made.)

Next, we tackle the Rayleigh example. 
Again comparing the fixed-order expression (the bottom line of \Eq{toy_FO_pdfs}) with \Eq{divide_out}, we see that $p_{m^*}(t) = t$, $m^* = 1$, and $p_m(t) = \frac{(-1)^{m-1}t^{2m-1}}{(m-1)!2^{m-1}}$ .
Once again, we implement \Eq{full_matching}:

\begin{align}
\begin{split}
g(t, \alpha) &= -\log(\alpha t) + [ \sum_{k}\frac{1}{k}\left(\left[-\sum_{m = 2}^M \frac{1}{(m-1)!}(\frac{-\alpha t^2}{2})^{m-1} \right]^k - \left[ \sum_{m=1}^{M-1} -\frac{1}{m!}(\frac{-\alpha t^2}{2})^m \right]^k\right) + \O\left(\alpha^{M}\right)]\\
 &= -\log(\alpha t) + [ \sum_{k}\frac{(-1)^k}{k}\left(\left[\sum_{m' = 1}^{M-1} \frac{1}{(m')!}(\frac{-\alpha t^2}{2})^{m'} \right]^k - \left[ \sum_{m=1}^{M-1} \frac{1}{m!}(\frac{-\alpha t^2}{2})^m \right]^k\right) + \O\left(\alpha^{M}\right)] \nonumber\\
 &= -\log(\alpha t) + \O(\alpha^{M}).
\end{split}
\end{align}
And so we extract 
\begin{align}
\begin{split}
&g^*(t,\alpha) = \alpha t\\
&g_\textrm{Analytic}(t, \alpha) = 0 + \O(\alpha^{M}). \label{eq:rayleigh_completion_g}
\end{split}
\end{align} 
Thus we can  completely specify the full RDF as
\begin{align}
\begin{split}
   q^{(\textrm{Rayleigh})(M)}(t | \alpha) &= \alpha t \exp\left( - \frac{\alpha t^2}{2}  \right)+ \mathcal{O}(\alpha^{M+1}). \label{eq:rayleigh_completion}
\end{split}
\end{align}
We are free to add any $\O(\alpha^M)$ contributions to $g_\textrm{Analytic}(t, \alpha)$, though again in this case choosing it to be zero happens to reproduce the true distribution.

While the two examples we have shown have happened to lead us to the all-orders PDFs when the higher-order terms are chosen to be zero, this behavior will not generically occur.
In particular, it is only when we have $m^*= 1$ and $\frac{p_{m+m^*}(t)}{p_{m^*}(t)} = \int_0^t dt' p_m(t')$ that the analytic matching procedure will return 0 for $g_{\rm Analytic}$ up to and including $\O(\alpha^{M-1})$.

The unfixed $g_m(t)$ are a type of ``theory uncertainty'', since in principle, we do not have any information as to what these are. 
We will discuss the interpretation of the unfixed $g_m(t)$ as theoretical nuisance parameters much more quantitatively and thoroughly in \Sec{nuisance_params}.\footnote{Why not now? Because in the RDF framework, theory uncertainties are due to proper nuisance parameters. That is, theory uncertainties are only meaningful if one is performing a statistical fit to data to infer some parameter. Here, we are just producing prediction curves without data.}
For now, we will randomly choose coefficients $g_{mn}$ as a proxy for qualitatively understanding this uncertainty --- formally, this is equivalent to placing a Bayesian prior on the higher-order terms.
A potentially natural choice is to choose $g_{mn} \sim \mathcal{N}(0, \frac{1}{m!n!})$, so that $g(t, \alpha) = \sum_{m, n} g_{mn} \alpha^m t^n$ converges reasonably quickly for all $\alpha, t$.
We also require that the highest power in $t$ is negative so that $g$ is bounded from above.
The hope is that this choice should envelope the ``true" answer, but this is not guaranteed. 
For simplicity, we will only take $g_{M+1}$ to be nonzero (that is, we take one higher order in $\alpha$), though we emphasize that this is just a simplifying choice within our ansatz.

\begin{figure}
    \centering
\includegraphics[width=0.49\linewidth]{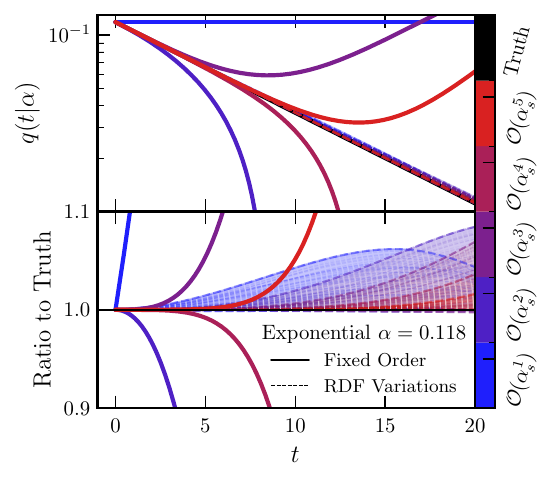}
\includegraphics[width=0.49\linewidth]{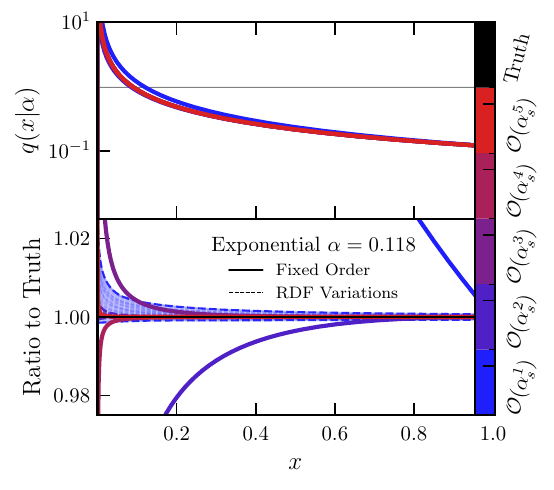}
    \caption{The RDF unitary completions of the exponential distribution, as given by \Eq{exponential_completion}, for random choices of the higher-order $g_m(t)$ parameters. The FO distributions are shown as thick colored lines, and the true distribution is shown as a black line. Each random choice of $g$ is shown as a thin dotted colored line. Note for $g = 0$, the completion lies exactly on top of the true distribution. To guide the eye, we draw envelopes around the random variations, though these envelopes are not themselves valid distributions. The distributions are shown as a function of $t$ (left) and $x$ (right).}
    \label{fig:exponential_completion}
\end{figure}

\begin{figure}
    \centering
\includegraphics[width=0.49\linewidth]{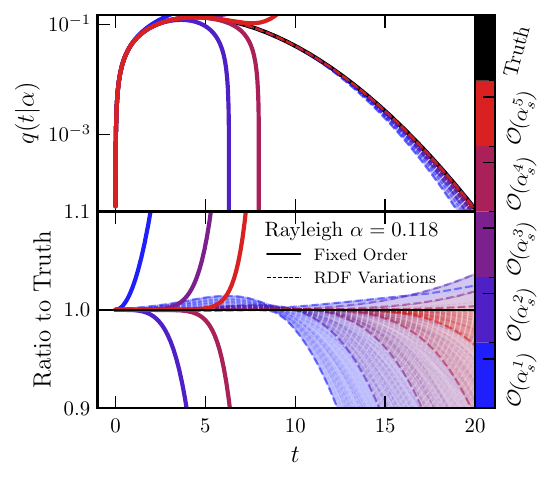}
\includegraphics[width=0.49\linewidth]{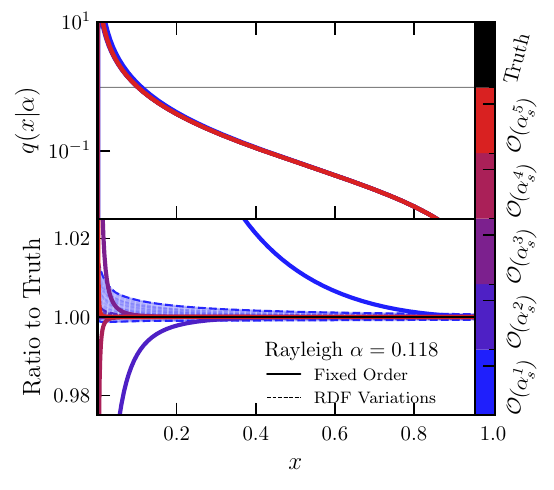}
    \caption{The same as \Fig{exponential_completion}, but for the Rayleigh distribution whose RDF completion is given by \Eq{rayleigh_completion}.}
    \label{fig:rayleigh_completion}
\end{figure}

In \Figs{exponential_completion}{rayleigh_completion}, we show unitary completions of the exponential distribution and Rayleigh distribution, as given by \Eq{exponential_completion} and \Eq{rayleigh_completion}, for random choices of the higher-order $g_m(t)$ parameters as described above.
Each thin line is a different choice of the higher-order terms of $g_M(t)$, with the color representing the order $M$ of the approximation.
Every single line here is positive, finite, and normalized, and it matches the perturbative expansion of the exponential up to and including $\mathcal{O}(\alpha^M)$.
When the higher-order terms are chosen to be zero, the RDF exactly reproduces the truth distribution.
We first note that, at least by eye, all possible choices track the true distribution, which indicates that this ``prior'' on $g$ is reasonable.
Second, we note that the distributions tighten as $M$ increases.
This reflects our greater degree of perturbative certainty: in the $M\to \infty$ limit, the FO expansion is exactly equal to the true distribution, and there is no extra freedom that can be accessed by varying higher-order coefficients.\footnote{The tightening is \emph{not} merely due to the factor of $\frac{1}{m!}$ in the prior --- it persists even when this is removed. }

\subsection{Numeric matching}
\label{sec:numeric_matching_toy}

We next show how the numeric matching procedure outlined in \Sec{numeric_matching} can be used to extract the functions $g^*(t, \alpha)$ and $g_m(t)$ for our two toy distributions.
For simplicity, we will assume these functions take the form of a power series in $t$ such that we are learning coefficients $g_{mn}$ attached to the terms $\alpha^m t^n$ in both $g^*$ and $g_{\rm Analytic}$.
 
Since the toy examples have known analytic FO PDFs given \Eq{toy_FO_pdfs}, we can generate ``idealized" histograms from the FO PDFs themselves, mimicking the output of a fixed-order program such as \texttt{EERAD3}~\cite{Aveleira:2025svg}, though with no MC uncertainties or other associated errors. 
We fix to the domain $t\in[0, 10]$ and define 200 evenly-spaced bins. 
Then, our ``histograms" come from evaluating the FO PDFs at the bin centers.
For all learnable objects, we use the reroll initialization procedure as outlined in \Sec{numeric_matching} --- for the Rayleigh distribution, the reroll initialization does nothing, as the original parameter choice of $g^*_{11} = 1$ is already near-optimal.
For each order of $\alpha$, we fit a polynomial in $t$ up to $t^7$ for each of the $g^*_{mn}$ and $g_{\text{Analytic}_{mn}}$ matrices).
Within each batch, we generate 320 choices of $\alpha$ uniformly in the interval [0.005, 0.325].
We take the bin error to be an arbitrary constant proportional to 1.0 for each bin, as the overall scale off the loss is irrelevant. 

\begin{figure}
    \centering
\includegraphics[width=0.435\linewidth]{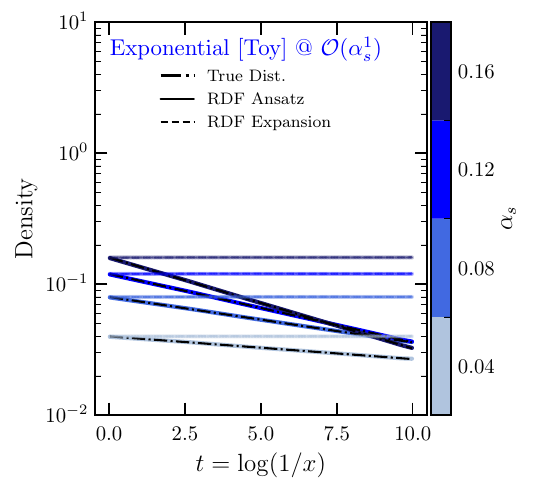}
\includegraphics[width=0.435\linewidth]{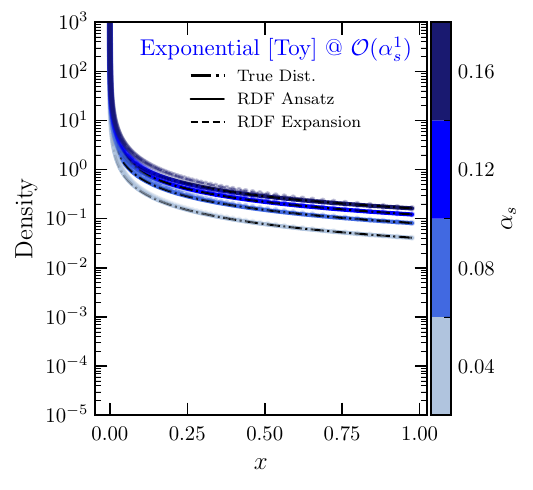}
\includegraphics[width=0.435\linewidth]{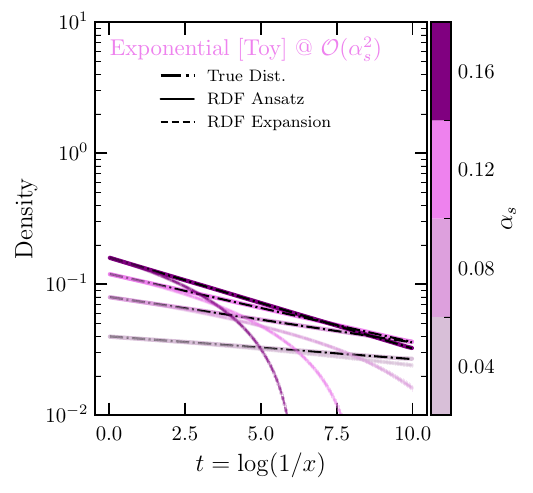}
\includegraphics[width=0.435\linewidth]{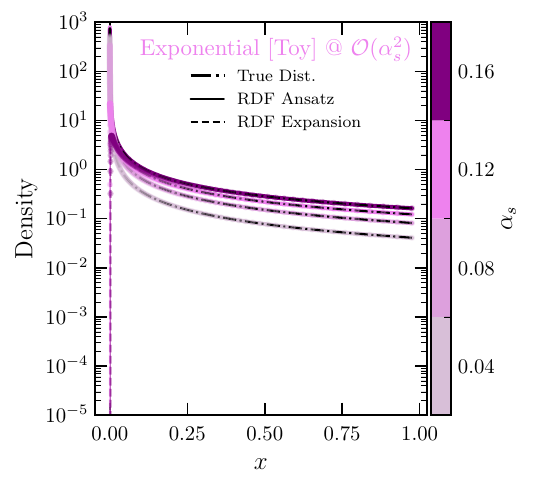}
\includegraphics[width=0.435\linewidth]{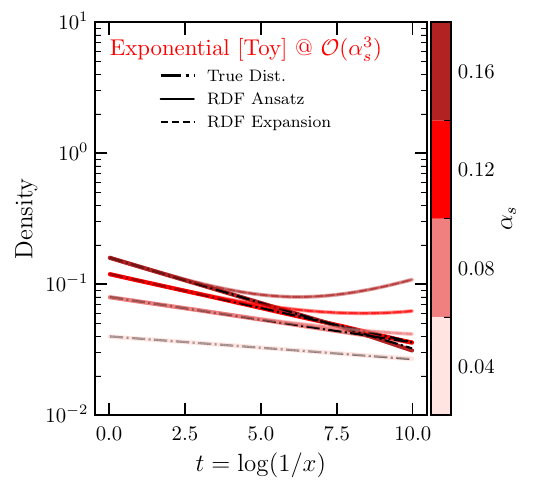}
\includegraphics[width=0.435\linewidth]{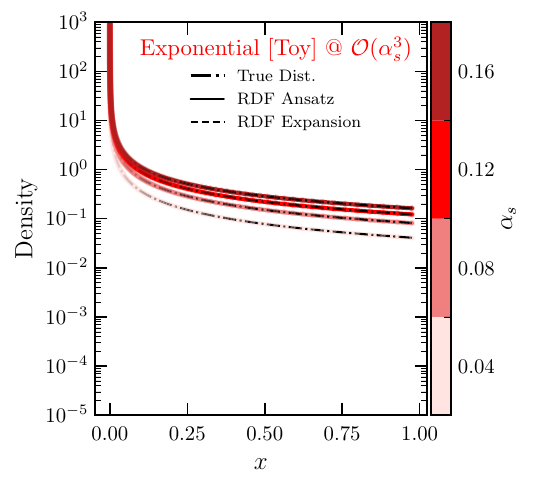}
    \caption{RDF numeric fits to the exponential toy at $\mathcal{O}(\alpha_s^1)$ (top), $\mathcal{O}(\alpha_s^2)$ (middle), and $\mathcal{O}(\alpha_s^3)$ (bottom), plotted as a function of $t$ (left) and  $x$ (right).
    For several values of $\alpha_s$, the
     RDF itself is shown as a solid line, and the Taylor expansion of the RDF is shown as a dotted line. 
    For comparison, we show the true exponential distribution as a black dash-dotted line.
    }
    \label{fig:exponential_matching}
\end{figure}

In \Figs{exponential_matching}{rayleigh_matching}, we show the results of numerically fitting to the exponential and Rayleigh toys at first, second, and third orders in $\alpha$.
All training hyperparameters are given in \Tab{hyperparams} in \App{training_hyperparams}.
In all cases, the Taylor-expanded RDF agrees exactly with the fixed-order target.
The good behavior of the RDF is especially visible when considering the $t$-space plots: while the fixed-order targets either diverge or go negative as $t \to \infty$, the RDF distributions simply tend to zero.
To give a taste of the numerics and minimization procedure, we show the learned values of the $g_{mn}$ matrices at each epoch of training for the $\O(\alpha_s^1)$ exponential in \App{supp_plots}.

\begin{figure}
    \centering
\includegraphics[width=0.435\linewidth]{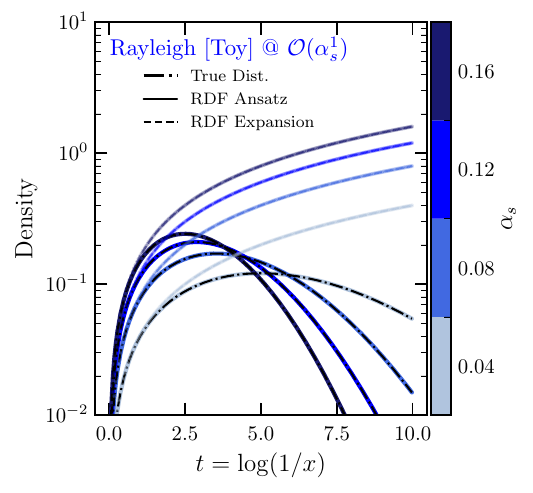}
\includegraphics[width=0.435\linewidth]{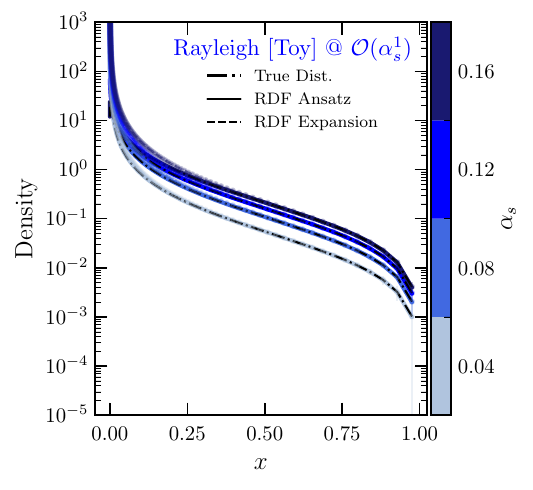}
\includegraphics[width=0.435\linewidth]{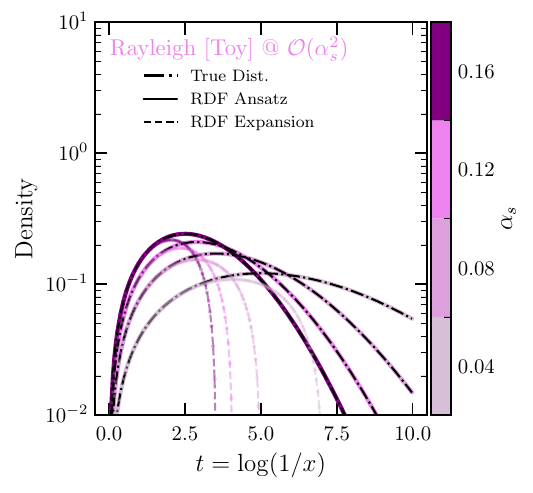}
\includegraphics[width=0.435\linewidth]{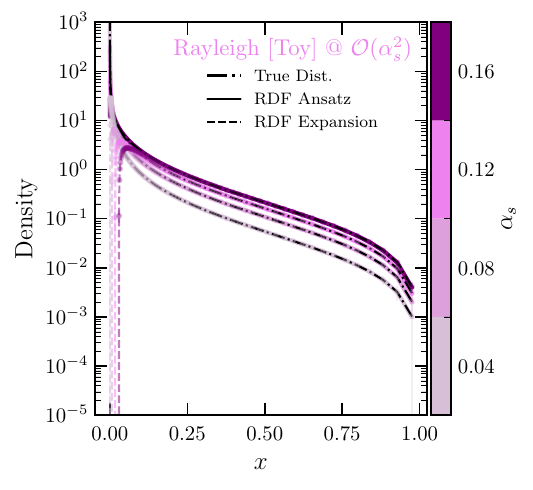}
\includegraphics[width=0.435\linewidth]{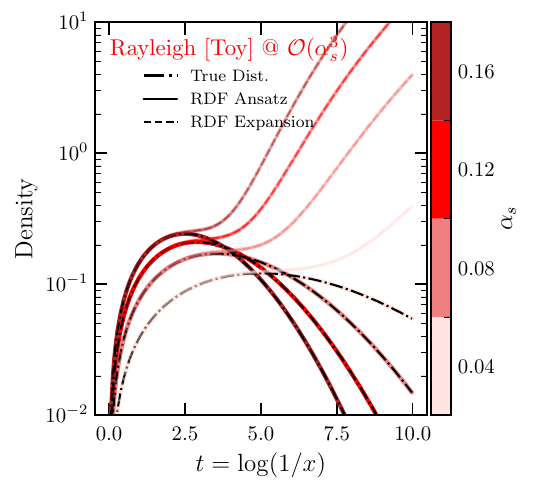}
\includegraphics[width=0.435\linewidth]{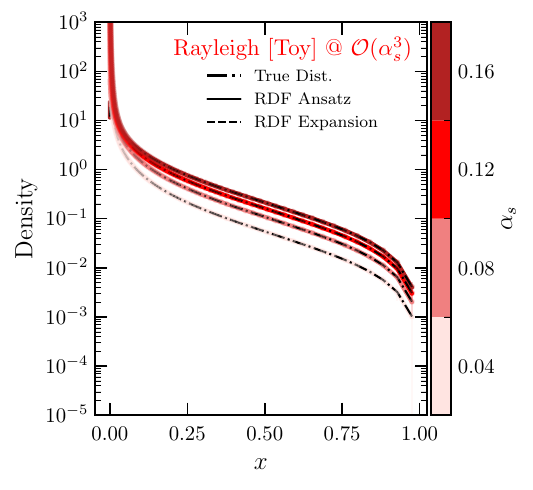}
    \caption{The same as \Fig{exponential_matching}, but for the Rayleigh toy rather than the exponential toy.
    }
    \label{fig:rayleigh_matching}
\end{figure}

\section{Matching to QCD Observables}
\label{sec:qcd_observables}

In this section, we explore applying the RDF to realistic QCD observables.
We first show how the RDF can match to \emph{jet angularities}~\cite{Berger:2003iw, Larkoski:2014uqa} for a single jet.
We then show how the RDF can work with simultaneous observables by analytically matching multiple angularities at once through $\O(\alpha_s^1)$.
We will see how the RDF can be used to complete fixed-order distributions with zero additional knowledge of QCD, and that while one is not guaranteed to recover the leading-logarithmic resummation, it is at least contained within the solution space.
%
%
We then numerically study the event thrust~\cite{Brandt:1964sa,Farhi:1977sg} through $\mathcal{O}(\alpha_s^3)$.
Here, no full fixed-order expressions are available beyond leading order, so we match the RDF to MC generated events.

\subsection{Jet Angularities}\label{sec:angularity}

Jet angularities $\lambda$ are a common observable for characterizing the angular ``spread'' of energy within a jet.
The Winner-Take-All (WTA) angularity is defined as:
\begin{align}
    \lambda^{(\beta)} = \sum_i z_i \left(\frac{\theta_i}{R}\right)^\beta,
\end{align}
where $\theta_i$ is measured relative to the WTA axis~\cite{Larkoski:2014uqa} of the jet, $R$ is the jet radius, and $\beta$ is an angular weighting exponent.
The WTA axis roughly aligns with that of the hardest parton in the jet.\footnote{More precisely, it is the axis corresponding to the hardest branch in a sequence of binary splittings.}
The full dynamic range of $\lambda$ is $\lambda \in [0, 1]$, but at any finite order in perturbation theory $M$, the range is $\lambda \in [0, 1 - \frac{1}{M+1}]$ due to the WTA axis selection.
For example, to first order, $\lambda \in [0, \frac{1}{2}]$.

To leading order in $\alpha_s$, we may calculate the differential distribution of the WTA angularity of a quark-initiated jet using the Altarelli-Parisi splitting function for a gluon emission off of a quark, $P(z) = \frac{1+z^2}{1-z}$~\cite{Altarelli:1977zs}:
\begin{align}
    p_{\rm FO}(\lambda) = \frac{\alpha_s C_F}{\pi} \int_0^R \frac{d\theta}{\theta} \int_0^1 dz\,  \left[  P(z) \,\delta\left(\lambda - \min(z, 1-z)\frac{\theta^\beta}{R^\beta}\right)\right].
\end{align}
Here, the $\min(z, 1-z)$ is due to the WTA condition for a single emission, as only the softer of the two particles contributes to the angularity.
We find:
\begin{align}
    p_{\rm FO}(\lambda) &= \left(\frac{\alpha_s C_F}{\pi\beta}\right) \frac{1}{\lambda}\left(2\log\frac{1-\lambda}{\lambda} + 3\lambda - \frac{3}{2}\right)\Theta\left(\frac{1}{2}-\lambda\right) \\
    &\downarrow \quad \lambda \to 0\qquad\text{(Soft and Collinear limit)}\nonumber\\
    &= \left(\frac{\alpha_s C_F}{\pi\beta}\right) \frac{1}{\lambda}\left(2\log\frac{1}{\lambda} \right).
\end{align}
Note that in the $\lambda \to 0$ (soft-collinear) limit, the $\Theta$-function vanishes.  
Defining:
\begin{align}
    t = \log{\frac{1}{\lambda}},
\end{align}
we may rewrite the above as:
\begin{align}
    p_{\rm FO}(t) &= \left(\frac{\alpha_s C_F}{\pi\beta}\right) \left(2t + 2\log(1 - e^{-t}) + 3e^{-t} - \frac{3}{2}\right)\Theta\left(t - \log2 \right) \label{eq:jet_angularity_t} \\
    &\downarrow  t \to \infty\qquad\text{(Soft and Collinear)}\nonumber\\
    &= \left(\frac{\alpha_s C_F}{\pi\beta}\right) \left(2 t\right)\label{eq:jet_angularity_t_soft_collinear}.
\end{align}
In $t$-space, to first order in $\alpha_s$ in the soft-collinear limit, the jet angularity in \Eq{jet_angularity_t_soft_collinear} takes exactly the same form as the first-order Rayleigh distribution of \Eq{toy_FO_pdfs}.
This should be expected, as the fixed-coupling double-logarithmic Sudakov approximation of jet angularities in the soft-collinear limit is exactly a Rayleigh distribution.

From \Eq{jet_angularity_t}, we can perform matching for the full jet angularity using the full analytic matching procedure in \Eq{full_matching}. 
Unlike the toy examples of \Sec{toy_examples}, however, we are no longer working with a simple polynomial in $t$.
We also now have nontrivial $\Theta$-functions enforcing kinematic boundaries on $\lambda$.
Following the analytic matching algorithm of \Eq{full_matching}, we may choose (with $m^* = 1$):
\begin{align}
    g(t, \alpha) = -\log\left[\left(\frac{\alpha_s C_F}{\pi\beta}\right) \left(2t + 2\log(1 - e^{-t}) + 3e^{-t} - \frac{3}{2}\right)\Theta\left(t - \log2 \right) + \O(\alpha_s^2) \right] + \O(\alpha_s^1), \label{eq:g_angularity}
\end{align}
where the two free higher-order $\O(\alpha_s^m)$ terms are constrained only by being analytic and bounded from above, corresponding to changing $g^*$ or $g_{\rm Analytic}$ respectively.
At first glance, it appears that this $g$ is worryingly non-analytic in $t$, due to the logarithms-of-logarithms in $t$.
However, this is fine because the requirement is that $g$ only has up to single-log nonalyticities in $\alpha_s$.
This choice leads to the RDF:
\begin{align}
    q(t |\alpha) &= \left[\left(\frac{\alpha_s C_F}{\pi\beta}\right) \left(2t + 2\log(1 - e^{-t}) + 3e^{-t} - \frac{3}{2}\right)\Theta\left(t - \log2 \right) \right] \nonumber \\
    &\quad\times \exp\left[-
\left(\frac{\alpha_s C_F}{\pi\beta}\right)
\left[
t^2 - (\log 2)^2
+ 2\left(\mathrm{Li}_2\left(e^{-t}\right)
      - \mathrm{Li}_2\left(\tfrac12\right)\right)
- 3\left(e^{-t} - \tfrac12\right)
- \frac{3}{2}\left(t - \log 2\right)
\right]\right]\label{eq:rdf_angularity}\\
&\downarrow  t \to \infty\qquad\text{(Soft and Collinear)}\nonumber\\
&= \left(\frac{\alpha_s C_F}{\pi\beta}\right) \left(2 t\right)  \exp\left(- \frac{\alpha_s C_F}{\pi\beta} t^2\right),\label{eq:rdf_angularity_lim}
\end{align}
where $\textrm{Li}_2$ is the dilogarithm function.
In the $t \to \infty$ limit, we successfully reproduce the double-log Sudakov factor result, which is precisely the Rayleigh distribution.
This is \emph{not} because the RDF knows about the all-orders emission structure of QCD (and therefore knows to reproduce the Sudakov factor due to factorized emissions), but rather because all the logarithms needed already appear at $\O(\alpha_s^1)$ and this is the only unitary way to combine them. 
We will see in the multivariate case in the next section that this full reproduction does not generally occur.

\begin{figure}
    \centering
\includegraphics[width=0.485\linewidth]{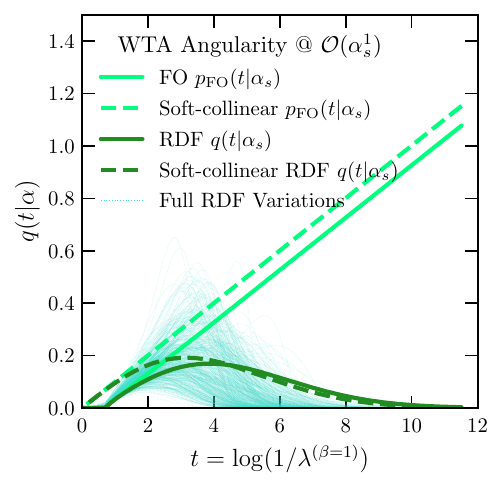}
\includegraphics[width=0.495\linewidth]{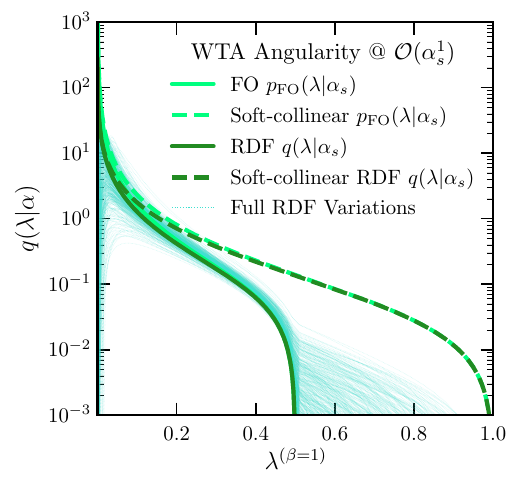}
    \caption{RDF analytic matching to the WTA jet angularity at $\mathcal{O}(\alpha_s^1)$, as given by \Eqs{rdf_angularity}{rdf_angularity_lim}, plotted as a function of $t$ (left) and the angularity $\lambda^{(\beta = 1)}$ (right). The original fixed-order expressions are shown in light green, and the RDF expressions are in dark green. The soft-collinear (double-log) limits are shown as dashed lines. Random variations of the higher-order terms for the full (i.e. \textit{not} soft-collinear RDF) of \Eq{g_angularity} are shown as thin turquoise lines.}
    \label{fig:angularity_o1}
\end{figure}

In \Fig{angularity_o1}, we show the RDF-matched angularity given by \Eqs{rdf_angularity}{rdf_angularity_lim} for $\beta = 1$.
As expected, the angularities resemble the $\O(\alpha_s^1)$ Rayleigh distribution of \Fig{rayleigh_completion}, with the key difference being the presence of the $\Theta$-functions due to the different dynamical range.
We can also visualize the effect of higher-order corrections to the full RDF by adding random $\O(\alpha_s^2)$ variations inside the logarithm and $\O(\alpha_s^1)$ variations outside.
As with the toys in \Sec{toys_analytic}, we choose these variations to be random polynomials in $t$, up to degree 4, with coefficients chosen according to $\mathcal{N}(0,\frac{1}{m!n!})$.
Unlike the toys, we also allow for a random $\Theta(t - \theta_m)$ multiplying each polynomial, where $0 < \theta_m < \log(2)$, to account for the fact that higher-order terms may increase the dynamic range of $t$.
The random variations in \Fig{angularity_o1} are qualitatively similar to the Rayleigh distribution variations of \Fig{rayleigh_completion} in that they tend to peak at a lower value of $t$ than the baseline distribution does, and they go to zero faster than the baseline does.
This behavior is expected, as a generic second-order correction typically has large negative contributions at large $t$, owing to the fact that many observables have alternating signs in their expansion due to negative exponentials.
Unique to the single jet angularity example, we see a small ``kink'' below $t = \log(2)$: below this point, only second-order terms can contribute.

\subsection{Multivariate Jet Angularities}\label{sec:multi_angularity}

We now consider a more complex application of the RDF: measuring \emph{two} angularities on a single jet: $\lambda_\alpha$ and $\lambda_\beta$ for $\alpha > \beta$, and using the RDF to complete the multi-dimensional distribution.
Note that since $\theta_i < R$, we have $\lambda_\alpha < \lambda_\beta$.
Working to order $\alpha_s^1$ and  entirely in the soft-collinear limit for simplicity, we can calculate the differential cross section~\cite{Larkoski:2013paa, Larkoski:2014tva}:
\begin{align}
    p_{\rm FO}(\lambda_\alpha, \lambda_\beta) = \frac{2\alpha_s C_F}{\pi (\alpha - \beta)}\frac{1}{\lambda_\alpha \lambda_\beta}\Theta(\lambda_\alpha^\beta - \lambda_\beta^\alpha)\Theta(\lambda_\beta - \lambda_\alpha).
\end{align}
Naively, this looks like it factorizes, but the $\Theta$ functions induce correlations and prevent a full factorization. 

We define $t_\alpha = \log(\frac{1}{\lambda_\alpha})$ and $t_\beta = \log(\frac{1}{\lambda_\beta})$.
Then, we have:
\begin{align}
    p_{\rm FO}(t_\alpha, t_\beta |\alpha_s) &= \frac{2\alpha_s C_F}{\pi (\alpha - \beta)}\Theta_{\alpha\beta} \\
    p_{\rm FO}(t_\beta | \alpha_s) &= \frac{2\alpha_s C_F}{\pi \beta} t_\beta \\
    p_{\rm FO}(t_\alpha | t_\beta, \alpha_s) &=  \frac{\beta}{\alpha - \beta} \frac{1}{t_\beta} \Theta_{\alpha \beta}.
\end{align}
For convenience, we have defined the symbol $\Theta_{\alpha\beta} = \Theta(t_\beta < t_\alpha < \frac{\alpha}{\beta} t_{\beta})$.
Now, we can do matching to the multidimensional RDF as given in \Eq{renormalizing_flow_multi}, by matching to each individual term sequentially.
We will choose $t_1 \equiv t_\beta$ and $t_2 \equiv t_\alpha$.
This choice of ordering \emph{does} affect the final RDF, but only at higher orders, and higher-order coefficients can always be chosen to cure this effect.\footnote{We could have chosen the other way around, or even arbitrary combinations of $t_\alpha$ and $t_\beta$. These choices will lead to the same results up to the given order, but with potentially different behavior for higher-order terms.}

For the random variable $t_{\beta}$, the problem reduces to just a single jet angularity, which in the soft-collinear limit reduces to a Rayleigh distribution as above in \Sec{angularity}.
Therefore, we can immediately say (working with $f = e^{-g}$ rather than $g$ for convenience):
\begin{align}
    f_\beta(t_\beta, \alpha_s) &= \frac{2 \alpha_s C_F}{\pi \beta} t_\beta e^{-g_{\beta}(t_\beta, \alpha_s)}, 
\end{align}
where $g_\beta$ is some analytic and bounded-from-above function that is at least $\mathcal{O}(\alpha_s)$.

The matching game is slightly more complicated for $f_\alpha(t_\alpha, t_\beta, \alpha_s)$ and $p(t_\alpha | t_\beta, \alpha_s)$, since they have no $\alpha_s$ dependence to leading order and nontrivial $\Theta$-function dependence.
Pushing through with $m^* = 0$, we find that
\begin{align}
    f_\alpha(t_\alpha, t_\beta, \alpha_s) &= \frac{\beta}{\alpha t_\beta - \beta t_\alpha }e^{-g_{\alpha}(t_\alpha, t_\beta, \alpha_s)}\Theta_{\alpha\beta}, \label{eq:f_alpha}
\end{align}
where, as usual,  $g_\alpha$ is another analytic and bounded-from-above function that is at least $\mathcal{O}(\alpha_s)$.
Importantly, the integrals of $f_\beta$ and $f_\alpha$ diverge with $t_\beta$ and $t_{\alpha}$ respectively, satisfying the normalization requirements for the RDF.

Then, plugging into the full multidimensional RDF (\Eq{renormalizing_flow_multi}), the solution is:
\begin{align}
    q(t_\alpha, t_\beta | \alpha_s) &= \frac{2\alpha_s C_F}{\pi (\alpha - \beta)}\left(1 + \O(\alpha_s)\right)e^{-\alpha_s\frac{C_F}{\pi \beta} t_\beta^2 - \mathcal{O}(\alpha_s)}\Theta_{\alpha\beta}.\label{eq:multi_angularity_answer}
\end{align}
In this expression, we do not include $g_{\alpha}$ and $g_{\beta}$, since their integrals are unknown, and we simply show the orders of the correction they imply.
Crucially, the undetermined part in the exponential of \Eq{multi_angularity_answer} is $\mathcal{O}(\alpha_s)$, \emph{not} $\mathcal{O}(\alpha_s^2)$.
This is because this undetermined part is dominated by $g_{\alpha}$.
That is, there are logarithms in $\lambda_{\alpha}$ and $\lambda_\beta$ that are unaccounted for.
We can contrast our result with the full LL calculation (reproduced from Eq. 3.4 of \Reference{Larkoski:2013paa})\footnote{Note that \Reference{Larkoski:2013paa} does not include the running of $\alpha_s$ in their LL calculation. For consistency, we will not either. As discussed in \Sec{resummation}, it is in principle possible to include effects due to running in the RDF, but we will not do this here.}, which contains extra terms in $t_\alpha$ that are only visible at $\O(\alpha_s^2)$:
\begin{align}
    q^{\rm LL}(t_\alpha, t_\beta |\alpha_s) &= \frac{2\alpha_s C_F}{\pi (\alpha - \beta)}\left(1 + \frac{2\alpha_s C_F}{\pi (\alpha - \beta)}\frac{(t_\beta - t_\alpha)(\beta t_\alpha - \alpha t_\beta)}{\beta}\right)e^{-\alpha_s\frac{C_F}{\pi}(\frac{t_\beta^2}{\beta} + \frac{(t_\alpha-t_\beta)^2}{\alpha-\beta})}\Theta_{\alpha\beta}.\label{eq:multi_LL}
\end{align}
While the RDF did indeed capture the LL-behavior of the toy observables in \Sec{toys_analytic} and single jet angularities in \Sec{angularity}, it is vital to emphasize that there is no guarantee that it must do so in general!
This multivariate jet angularities example gives us one situation where the RDF did not capture the full LL result.
The RDF \emph{only} uses information at the given order in perturbation theory, while a genuine logarithmically resummed calculation requires additional information (e.g. factorized soft/collinear emissions, Sudakov structure, etc), all of which are assumptions about the higher-order structure of QCD.
Note, however, that this does not mean that we cannot include this information inside the RDF ansatz, if we so choose.
By strategically choosing the $\O(\alpha_s^1)$ function $g_\alpha$ in \Eq{f_alpha} such that
\begin{align}
    f^{\rm LL}_\alpha &= \frac{\beta}{\alpha t_\beta - \beta t_\alpha  }\Theta_{\alpha\beta} + \frac{2\alpha_sC_F}{\pi (\alpha-\beta)}(t_\alpha-t_\beta)\Theta_{\alpha \beta}\label{eq:oracle_f},
\end{align}
which merely involves picking the higher-order terms in $g^* \in g$ to be polynomial in $t_\alpha$ and $t_\beta$, we reproduce the LL calculation.
Here, we stress that we are working in a fixed-coupling approximation.
The point to emphasize is that while the RDF \emph{can} parameterize the logarithmic resummation, without prior knowledge or additional asummptions about the structure of the theory, one would not know to write \Eq{oracle_f} using only the information from a leading-order calculation.
Indeed, we do match the LL calculation to $\O(\alpha_s^1)$.
The stray terms are only visible at $\O(\alpha_s^2)$, and thus they \emph{would} be captured had we matched to an $\O(\alpha_s^2)$ fixed-order calculation.

\begin{figure}[tbh]
    \centering
    \includegraphics[width=0.75\linewidth]{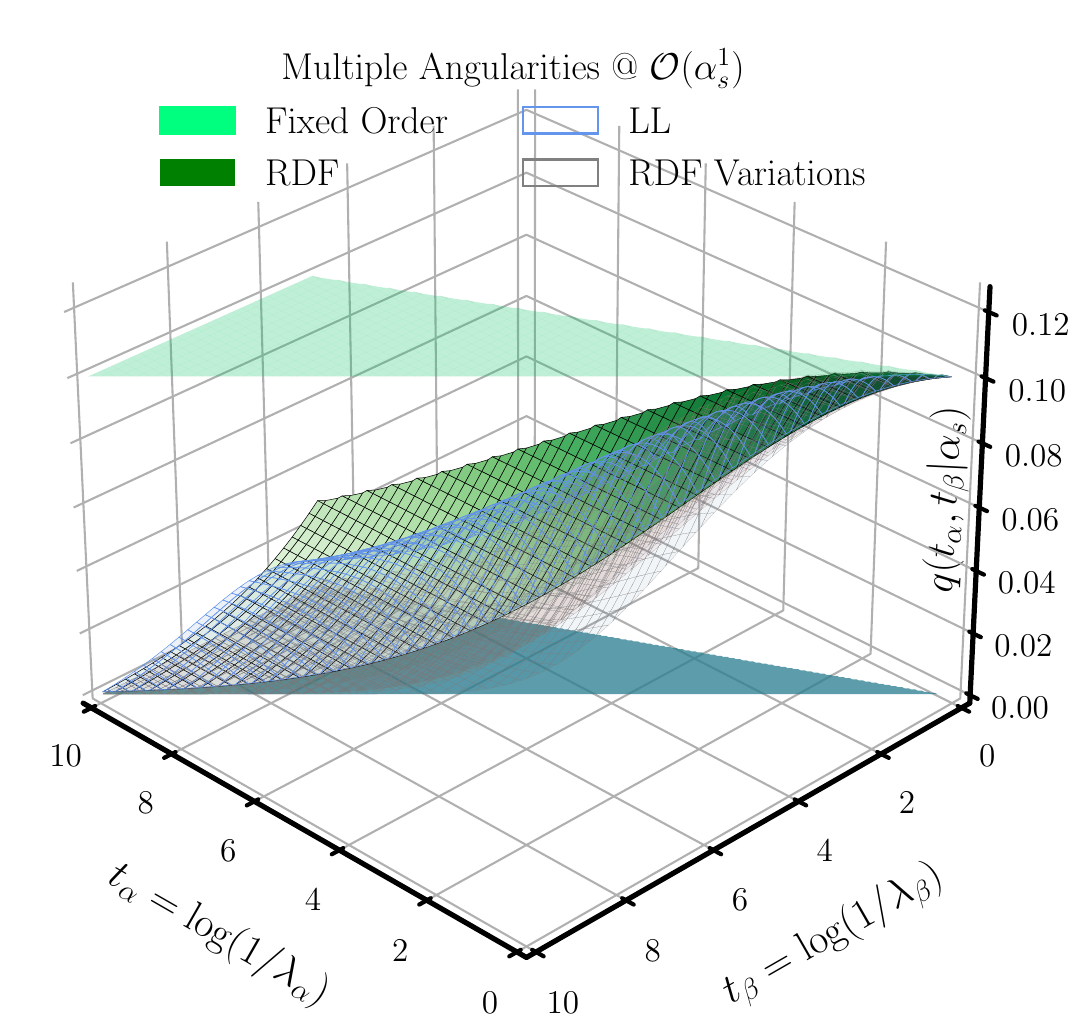}
    \caption{RDF analytic matching to the simultaneous jet angularities at $\mathcal{O}(\alpha_s^1)$, as given by \Eq{multi_angularity_answer}, plotted as a function of $t_{\alpha = 2}$ and $t_{\beta = 1}$. The original fixed-order expression is shown in light green, and the RDF expression is in dark green. The LL calculation~\cite{Larkoski:2013paa} from \Eq{multi_LL} is shown as a blue surface. Random variations of the higher-order terms of \Eq{multi_angularity_answer} are shown as thin gray surfaces. A shadow is shown on the $q = 0$ plane to indicate the domain of the function, $t_{\beta} < t_{\alpha} < \frac{\alpha}{\beta} t_\beta$.}
    \label{fig:multi_angularity}
\end{figure}

In \Fig{multi_angularity}, we show the RDF multi-angularity given by \Eq{multi_angularity_answer} as well as the usual higher-order random variations. 
Here, we only vary $g_{\alpha}$ and not $g_{\beta}$, as the latter is subleading.
The variations are the same as those in \Sec{angularity}, though now we have 2D polynomials rather than 1D.
This figure is the higher-dimensional analogue of \Fig{angularity_o1} (more precisely, \Fig{angularity_o1} is the marginal over $t_{\alpha}$ of \Fig{multi_angularity}). 
The RDF is able to properly normalize the fixed-order distribution while still respecting all kinematic boundaries. 
For comparison, the full LL result (\Eq{multi_LL}) is also shown.
The LL result is captured by variations of $g_{\alpha}$, as expected.
Visually, the LL result and the RDF result are similar.

As a cross-check, we can compute the marginals of \Eq{multi_angularity_answer} with respect to $t_\alpha$ and $t_\beta$ respectively to verify that we reproduce the expected single-angularity results of \Sec{angularity}.
They are:
\begin{align}
    q(t_\beta | \alpha_s) &= \frac{2\alpha_s C_F}{\pi \beta} t_\beta e^{-\frac{\alpha_s C_F}{\pi \beta} t_\beta^2}\\
    q(t_\alpha | \alpha_s) &= \frac{2\alpha_s C_F}{\pi (\alpha-\beta)}\frac{\sqrt{\pi}}{2\sqrt{\frac{\alpha_s C_F}{\pi \beta}}}\left[\text{erf}\left(\sqrt{\frac{\alpha_s C_F}{\pi \beta}} t_\alpha\right) - \text{erf}\left(\frac{\beta}{\alpha}\sqrt{\frac{\alpha_s C_F}{\pi \beta}} t_\alpha\right)\right] .
\end{align}
While we get the expected leading-log form for the $t_\beta$ marginal, we do not for the $t_\alpha$ marginal (though both at least have the correct fixed-order $\alpha_s\to 0$ limit).
One should not lose hope, however, as the usual Sudakov factor is hiding inside --- it is just obscured by the presence of higher-order $g$ terms that are allowed to be ignored to the order we are working in.
By expanding the error functions, we can derive the $g$-function corresponding to $q(t_\alpha | \alpha_s)$: 
\begin{align}
    q(t_\alpha | \alpha_s) \approx \text{RDF corresponding to } g = -\log(\frac{2\alpha_s C_F}{\pi \alpha} t_\alpha)  + \frac{\alpha_s C_F}{\pi \beta}\frac{t_\alpha^3}{3}\left(1 + \frac{\beta}{\alpha} + \frac{\beta^2}{\alpha^2}\right) + \mathcal{O}(\alpha_s^2). 
\end{align}
The first term is responsible for the expected leading-log form.
The second term is then the culprit --- since $m^* = 1$, the RDF algorithm specified in \Sec{analytic_matching} allows us to completely ignore these terms if we wish.
Thus, the multivariate RDF really does produce the correct marginals, albeit with some higher-order decoration that we can choose to ignore.

\subsection{Thrust: Numeric Matching up to $\O(\alpha_s^3)$}\label{sec:thrust_numeric}

In this section, we apply the numeric matching procedure from \Sec{numeric_matching} to event thrust~\cite{Brandt:1964sa,Farhi:1977sg} to obtain RDF predictions up to $\mathcal{O}(\alpha_s^3)$.
To accomplish this, we will use \texttt{EERAD3}~\cite{Aveleira:2025svg} to obtain numeric estimates of the fixed-order calculation before applying numeric RDF matching as was done for the toys in \Sec{numeric_matching_toy}.

Event thrust $\tau = 1-T$ measures how ``pencil-like'' an event is.
It is defined as:
\begin{align}
    T = \max_{\Vec{n}}\frac{\sum_i \Vec{p_i} \cdot \Vec{n}}{\sum_i |\Vec{p_i}|},
\end{align}
 where $i$ indexes particles, $\Vec{p_i}$ is the 3-momentum of that particle in the center-of-mass frame, and $\Vec{n}$ is the ``thrust axis''.
 The thrust $\tau$ takes on values between 0, for perfectly back-to-back events, to 0.5, for perfectly isotropic events.
 We elect to work with $x = 2\tau$ and $t = \log(1/2\tau)$ when interfacing with the RDF, as it is simplest when the dynamic range of the random variable is between 0 to 1 (in $x$ space) or 0 to $\infty$ (in $t$ space).
At any finite order in perturbation theory, however, the dynamic range of thrust is limited due to kinematics: in particular, at leading order, $\tau$ only ranges from 0 to $\frac{1}{3}$.

To obtain fixed-order numeric calculations of thrust in $e^+e^-$ collisions, we use \texttt{EERAD3} (Version 2, which we denote \texttt{EERAD3v2})~\cite{Aveleira:2025svg}, a public package for computing observables to up to $\O(\alpha_s^3)$~\cite{Gehrmann-DeRidder:2007nzq, Gehrmann-DeRidder:2007vsv}.
For an event shape observable $x$, \texttt{EERAD3v2} will compute functions $A(x)$, $B(x)$, and $C(x)$ such that the differential cross-section is
\begin{align}
    \frac{1}{\sigma}\dv{\sigma}{x} = \left(\frac{\alpha_s(\sqrt{s})}{2\pi}\right)\dv{A(x)}{x} + \left(\frac{\alpha_s(\sqrt{s})}{2\pi}\right)^2\dv{B(x)}{x} + \left(\frac{\alpha_s(\sqrt{s})}{2\pi}\right)^3\dv{C(x)}{x} + \O(\alpha_s^4)
\end{align}
at a fixed renormalization scale $\mu = \sqrt{s}$.
These coefficients can then be used to build histograms of the distribution of $x$ for various values of $\alpha_s(\sqrt{s})$.

We choose $\sqrt{s} = m_Z \approx 91.2$ GeV and simulate $Z \to q\bar{q}$ (\texttt{EERAD3v2} process ID \texttt{1}) with 3 hard jets, so that $\O(\alpha_s^1)$ corresponds to LO, $\O(\alpha_s^2)$ corresponds to NLO, and $\O(\alpha_s^3)$ corresponds to NNLO.
We use $10^8$ phase space points (100k shots per run across 1000 runs with different seeds) for the LO and NLO\footnote{This took about 24 hours on a CPU cluster.}, and NNLO\footnote{This took about 3 weeks on the same CPU cluster.}.
All other settings are left as default.
The end results are the $A$, $B$, and $C$ functions from which observable histograms can be built.

We then generate the distributions of the thrust $\tau = 1-T$ using \texttt{EERAD3v2}'s \texttt{makedist} command.
For a fixed value of $\alpha_s(m_Z)$, we may generate histograms for $\log(\tau)$.
Histograms are generated such that $\log(\tau)$ is between -10 and 0, with 200 uniform bins. 
The kinematically-disallowed values of the thrust histogram have bin values of zero.
We repeat this for 320 values of $\alpha_s(m_Z)$, uniformly chosen between 0.005 and 0.325.
\texttt{EERAD3v2} also reports a Monte-Carlo uncertainty and a ``theory-uncertainty'' (in quotations, since we will define a new theory uncertainty in \Sec{nuisance_params}) due to scale variations on each bin.
We include both in quadrature as the total error used in the MSE loss \Eq{numeric_mse_loss}.
By default, the error reported on the kinematically-disallowed bins is zero.
Rather than removing these bins entirely (as it is important to reproduce that these bins are empty), we ``clip'' all errors from below, such that zero-errors are replaced by the error on the first nonzero bin.
We emphasize that this choice is just as a numeric regulator to avoid divide-by-zero errors in \Eq{numeric_mse_loss} --- in principle, if one could perfectly guess the form of all $\Theta$-functions in the ansatz without gradient descent, this regularization would not be necessary as the loss would be finite.

\begin{figure}
    \centering
\includegraphics[width=0.435\linewidth]{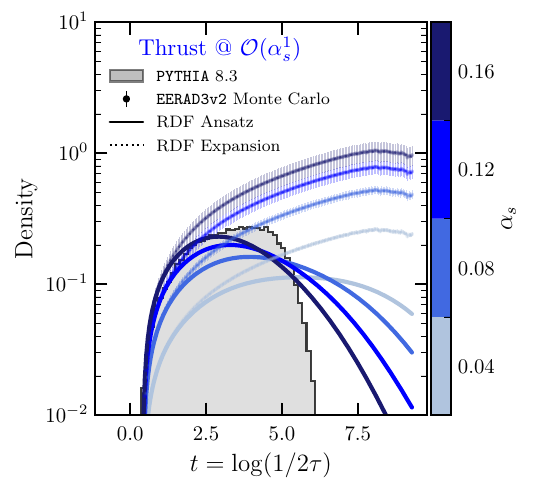}
\includegraphics[width=0.435\linewidth]{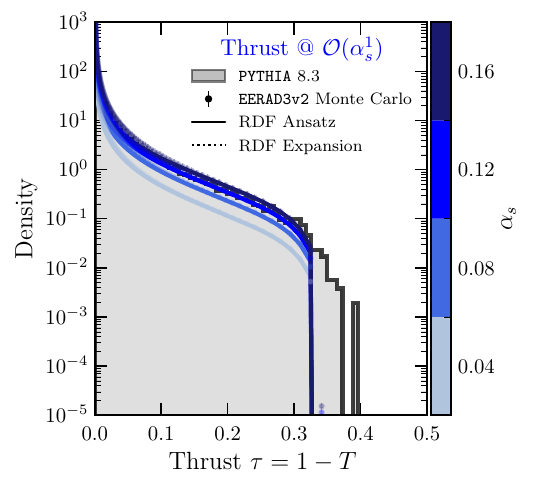}
\includegraphics[width=0.435\linewidth]{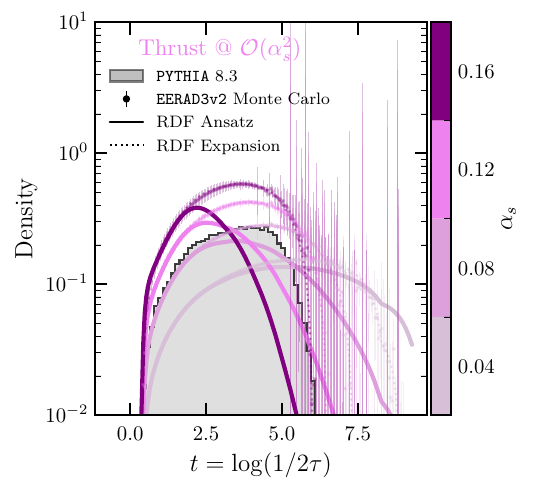}
\includegraphics[width=0.435\linewidth]{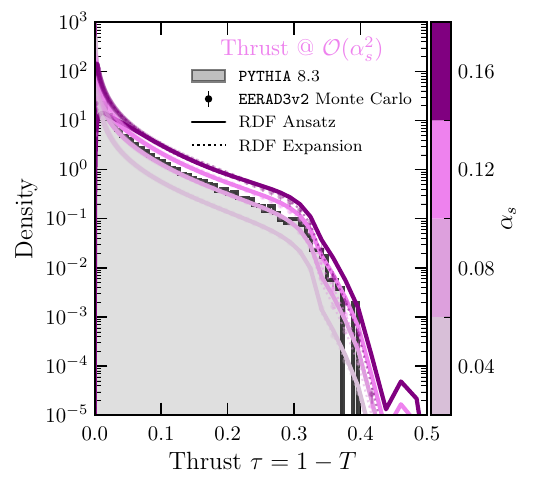}
\includegraphics[width=0.435\linewidth]{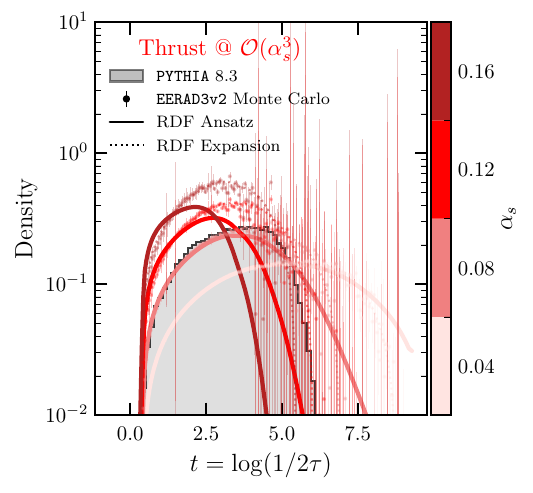}
\includegraphics[width=0.435\linewidth]{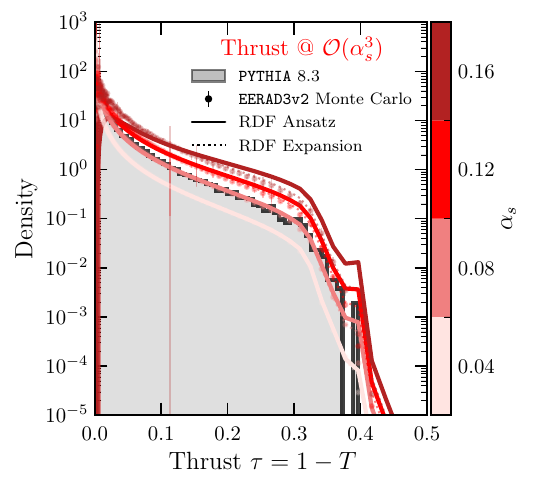}
    \caption{RDF numeric fits to event thrust at $\mathcal{O}(\alpha_s^1)$ (top), $\mathcal{O}(\alpha_s^2)$ (middle), and $\mathcal{O}(\alpha_s^3)$ (bottom), plotted as a function of $t$ (left) and thrust $\tau$ (right).
    For several values of $\alpha_s$,
    the original MC distribution calculated using $\texttt{EERAD3v2}$ are shown as points with error bars. The RDF itself is shown as a solid line, and the Taylor expansion of the RDF is a dotted line, which ideally should match the MC calculation. 
    For comparison, we also show a calculation of the event thrust with $\textsc{Pythia}$ 8.3.
    }
    \label{fig:thrust_matching}
\end{figure}

In \Fig{thrust_matching}, we show the results of numerically matching the event thrust at first, second, and third orders in $\alpha_s$.
All training hyperparameters are given in \Tab{hyperparams}.
For comparison, we also show the thrust as generated with \textsc{Pythia} 8.3~\cite{Sjostrand:2014zea} in $e^+ e^- \to q\bar{q}$ events at the $Z$-pole.
For all three orders, the Taylor-expanded ansatz sits almost perfectly on top of the \texttt{EERAD3v2} target. 
However, there is a notable discrepancy between the $\alpha=0.12$ RDF ansatzes and the \textsc{Pythia} prediction at the default tune.\footnote{The default Monash 2013 tune of \textsc{Pythia}~\cite{Skands:2014pea} uses $\alpha_s = 0.1365$ as its FSR showering parameter. This differs from the expected $0.118$, but the \textsc{Pythia} value is based on the CMW~\cite{Catani:1990rr} scheme rather than $\overline{\text{MS}}$, and is also different due to tuning.}
In particular, at order 1, the RDF ansatz tends to underfill the thrust phase space: the maximum thrust is $\sim 0.35$ compared to the \textsc{Pythia} maximum of 0.4. 
At orders 2 and 3, the RDF correctly drops off quickly near this kinematic threshold, though it is not sharp because of the $T$ parameters in the numeric RDF.
However, we should not expect the \textsc{Pythia} and the RDF predictions to completely agree: the former is an LL calculation including additional effects such as hadronization (which is known to shift the distribution), kinematic conservation, etc, while the latter is simply a unitary extension of a fixed order calculation.

\section{Nuisance Parameters and Theory Uncertainties}\label{sec:nuisance_params}

Having matched to fixed order calculations in \Sec{qcd_observables}, in this section we explore how the higher-order parameters of the RDF can be used as nuisance parameters to capture perturbative uncertainties when fitting to data.
These nuisance parameters capture the uncertainty due to the higher-order terms being unknown, akin to \Reference{Tackmann:2024kci}.
We will perform fits of $\alpha_s$ using up to the $\O(\alpha_s^3)$ matched RDFs using publicly-available event shape data from ALEPH~\cite{ALEPH:2003obs, hepdata.12794.v1/t54}.

Our approach is simple: given that we have a $g$-function matched to order $M$, we simply add additional orders using the same parametric form given in \Eq{numeric_ansatz}.
That is, we write:
\begin{align}
    g(t, \alpha) = \underbrace{g_{\rm Matched}(t, \alpha, \phi)}_{\mathcal{O}(\alpha^M_s)} + \underbrace{g_{\rm Nuisance}(t, \alpha; \nu)}_{\mathcal{O}(\alpha^{M+1}_s)},
\end{align}
where $\phi$ represents the seven already-matched components of $g$ ($g^*_{mn}$, ${g_\mathrm{Analytic}}_{mn}$, $\theta^*_m$, ${\theta_\mathrm{Analytic}}_m$, $T^*$, $T_\mathrm{Analytic}$, and $T_\textrm{abs}$), and $\nu$ are nuisance parameters representing higher-order corrections in $\alpha_s$ to $g$.
To illustrate explicitly: if we consider the array $g^*_{mn}$, rows $m^*$ through $M$ (inclusive) are contained in $\phi$ (as they would be determined through the numeric matching procedure), and row $M+1$ would be contained in $\nu$.
Then, given data $t_i$, we can perform a likelihood fit (either a full likelihood if event-level data is available, or a binned likelihood) simultaneously for $\alpha$ and $\nu$ and profile over $\nu$, which we will discuss more in \Sec{fitting}.

For this procedure, we \textit{freeze} the components of $g_{\rm Matched}$ determined during numeric matching; only the $g_{\rm Nuisance}$ are varied in the fit.\footnote{The nuisance parameters probe one order of $\alpha_s$ higher than the corresponding matching procedure does. We will always refer to the RDFs as being of order $\mathcal{O}(\alpha_s^{M})$, where $M$ was the order of the matching calculation. For example, an $\mathcal{O}(\alpha_s^{3})$ RDF has nuisance parameters of order $\mathcal{O}(\alpha_s^{4})$.}
The likelihood requires an additional ``prior'' or regulator term for $\nu$ that sets a canonical scale for each parameter, which we will discuss more in \Sec{priors}.

\subsection{Fitting procedure}\label{sec:fitting}

We use a standard profile likelihood minimization procedure to extract $\alpha$ from experimental data.
We define a likelihood function between the RDF and the (binned) data as\footnote{We do not write the likelihood $\mathcal{L}$ as being a function of $\phi$, since $\phi$ is frozen to its values from the matching procedure.}
\begin{align}
\begin{split}
-\log\mathcal{L}(\alpha, \nu) = \frac{1}{2}\sum_{\textrm{Bin}_i}\frac{|\textrm{RDF}(\textrm{Bin}_i, \alpha, \phi, \nu)- \textrm{data}(\textrm{Bin}_i)|^2}{\textrm{error}(\textrm{Bin}_i)^2} + \frac{1}{2}\sum_k\frac{|\nu_k - \mu_k|^2}{\sigma_k^2}.
\label{eq:log_likelihood}
\end{split}
\end{align}

We assume Gaussian likelihoods for both the observable bin counts and the nuisance parameters, which is typically a good approximation in the limit of large statistics.
In principle, one could use the full event likelihood for the first term ($\sum_{i \in \text{Data}} \log(q(t_i |\alpha_i, \phi, \nu))$) since $q$ is by construction a probability distribution, though collider data is often  not made public in this format.
Likewise, the prior on parameters need not be Gaussian; this is simply a choice.\footnote{We also tried lognormal priors with a variety of means and standard deviations, but they provided very similar results to the Gaussian priors.}
We take $\mu_k = 0$ and $\sigma_k = \frac{\sigma}{m!n!} $ for the two $g$ matrices, where $\sigma$ is some $\O(1)$ scale that we are free to choose.
For the $T$ and $\theta$ parameters, we choose not to include a prior term (equivalent to choosing a flat unnormalized prior).
We do not claim that these particular prior choices are optimal, and we will discuss priors more thoroughly in the following subsection.
We stress that the MSE is taken between the non-Taylor expanded RDF and data.
This is as opposed to in the numeric matching procedure to determine $\phi$, where we take an MSE between the Taylor-expanded RDF and a potentially non-unitary target.

We define the profile likelihood function between the RDF and the data to fit as 
\begin{align}
\begin{split}
\log \mathcal{L}(\alpha) = \log\frac{\mathcal{L}(\alpha, \hat{\nu})}{\mathcal{L(\hat{\hat{\alpha}}, \hat{\hat{\nu}}})}, \label{eq:profile}
\end{split}
\end{align}
where we take the ratio of the log likelihood function minimized over the nuisance parameters to the log likelihood function minimized simultaneously over $\alpha$ and the nuisance parameters. 
The minimization is done numerically, similar to the numeric matching described in \Sec{numeric_matching}.
To initialize the nuisance parameters, we use the reroll initialization procedure outlined in \App{initialization}, except we change the initial variance from 0.1 to 1.
For numeric convenience, we scale the nuisance parameters $g_{mn} \in \nu$ by a factor of $\alpha^m$, so we really fit $g' = \alpha^m g$ in such a way that \Eq{profile} is invariant.
This allows for approximately the same scalings for the $\nu$ parameters as a function of $\alpha$, which makes the fitting easier.
For the toy examples, we fit to pseudodata\footnote{We call it ``pseudodata'' to emphasize that unlike the thrust example, this is just a toy and not real data from a real experimental collaboration.} generated according to the true all-orders distributions.
For the thrust, we fit directly to ALEPH data.
More details of the fitting procedure, especially as it differs from the matching procedure numerically, can be found in \App{min}.

Once we have constructed the profile likelihood function, we extract the best-fit $\alpha$ by scanning over the profile likelihood ratio and finding the value of $\alpha$ that minimizes it. 
We may also construct $1\sigma$ confidence intervals on $\alpha$ where $-2\log \mathcal{L}(\alpha) < 1$.
In doing this, we are implicitly assuming that we are in the region of validity of Wilks' Theorem \cite{Wilks:1938dza}.
However, we have not performed any quantitative study to test the coverage of these confidence intervals.

\subsubsection{Priors}\label{sec:priors}

While a feature of the RDF is that it parameterizes all possible consistent higher-order terms, this is both a blessing and a curse.
If one truly has no knowledge of higher-order coefficients, then it is impossible to meaningfully assign a finite theory uncertainty on $\alpha_s$, as there always exists some choice of higher-order terms that can absorb any change in $\alpha_s$.
For example, in the exponential toy, we have $f(t, \alpha) = \alpha + \O(\alpha^2)$. 
If $\O(\alpha^2)$ is infinitely flexible (even within the unitarity constraint), we can always write $\O(\alpha^2) = \nu_2 \alpha^2$ it is possible for the nuisance parameter $\nu_2$ to equal $\frac{c}{\alpha_0}$, where $\alpha_0$ is a constant that happens to have the same value as the frozen value $\alpha$, such that $f = (1 + c)\alpha$. 
This occurs when the likelihood has a flat direction in the $\nu_2$-$\alpha$ plane.
It is clear then that for any value of $\alpha$, there is a choice of nuisance parameter $c$ that will give the exact same function, essentially destroying all information in $\alpha$.
This example illustrates the need to have \emph{some} type of prior on higher-order terms, such as a constraint on the terms allowed in $\O(\alpha^2)$ (which can be achieved by scale variations, or by explicit choice as in~\Reference{Tackmann:2024kci}, or by regulating the size of the nuisance parameters by adding explicit ``prior'' terms to the likelihood).
Note that ``scale variations'' are indeed a prior: the common choice to vary scales by a factor of 2 is largely arbitrary.

We consider this need to choose a prior to be an essential feature rather than a detriment.
Theoretical uncertainties are \emph{not} statistical quantities,\footnote{More precisely, they are not frequentist in the sense that one cannot do repeated independent ``experiments'' to obtain a statistical estimator for what the true terms are. One must simply posit some prior for the allowed terms.} and thus it is only sensible to discuss them in a Bayesian setting. 
The RDF framework does not return any information about $\alpha_s$ when the prior is taken to be completely flat and higher-order terms are allowed to be anything (even when constrained by unitarity).
This behavior is expected and desired, and we show an example of it in \Fig{thrust_fit_inf} of \App{other_priors}, where an infinitely flat prior\footnote{Here, we consider the ``flat prior'' to be the same as ``no prior'', as a a genuinely flat prior over all possible values of $g_{mn}$ is not a normalized distribution. Any properly normalized prior over $g$ will necessarily contain some restriction on the scale of $g$.} has ``infinitely wide'' confidence intervals for $\alpha_s$.\footnote{Technically, even the choice of how many and which terms to include is also a prior. Just by choosing to only allow our nuisance parameters to be up to one higher order and using the same polynomial order, we have picked an extremely mild prior.}
The key takeaway is that priors on theory uncertainties are a desirable  feature.
On the other hand, we can easily place ``reasonable" priors on the RDF nuisance parameters to constrain higher-order information.
By ``reasonable", we simply mean that small changes to the functional form of the prior do not lead to large changes in downstream results.
We will show examples of this in the rest of this section.

\subsection{Example: Fits to toys}
As a warm-up, we first apply the fitting procedure to the toy examples introduced in \Sec{toy_examples}, i.e. the exponential and Rayleigh distributions. 
This subsection may be skipped by readers primarily interested in $\alpha_s$ extractions from real data.

\begin{figure}[tbh]
    \centering
\includegraphics[width=0.495\linewidth]{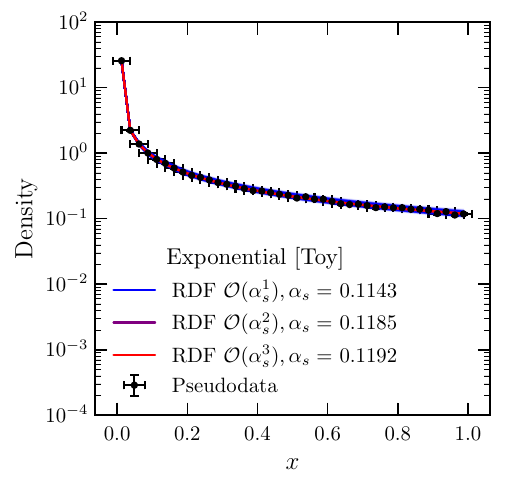}
\includegraphics[width=0.495\linewidth]{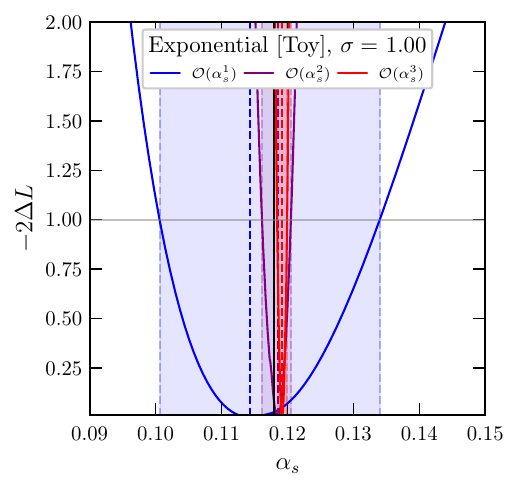}
    \caption{ (Left) RDF fits to exponential toy pseudodata with a true value of $\alpha = 0.118$. For each of $\O(\alpha_s^{1})$ (blue), $\O(\alpha_s^{2})$ (purple), and $\O(\alpha_s^{3})$ (red), the best-fit RDF is plotted as determined by the fitting procedure for the $\sigma = 1.0$ prior. (Right) The profile-log-likelihood, as defined in \Eq{log_likelihood}, of each of the three orders as a function of $\alpha_s$ for the $\sigma = 1.0$ prior. The minima are indicated by vertical lines, and confidence intervals are drawn where $-2\Delta L = 1$. The true value ($\alpha_s = 0.118$) is indicated by a vertical black line.
    }
    \label{fig:exponential_fit}
\end{figure}

To construct the ``data" used in the log likelihood function from \Eq{log_likelihood}, we generate histograms from the all-orders expressions given in \Eq{toys_closed_form}. 
We use 250,000 samples over 40 bins in $x$-space, ranging from 0 to 1, and we assign statistical errors of $\sqrt{N_i}$ to each bin.
In \Fig{exponential_fit}, we show the results of extracting $\alpha_s$ to first, second, and third orders on the exponential toy.
In the left panel, we show the ``experimental pseudodata" corresponding to an exponential distribution and the RDFs at orders $\mathcal{O}(\alpha_s^{1})$, $\mathcal{O}(\alpha_s^{2})$, and $\mathcal{O}(\alpha_s^{3})$ --- that is to say, the RDFs matched to first, second, and third order in $\alpha_s$ from \Sec{numeric_matching_toy}, then minimized over one higher order in the nuisance parameter.
In the right panel, we show the profile likelihood as a function of $\alpha$ using a Gaussian prior with $\sigma=1.0$ for the nuisance parameters.
The recovery of $\alpha_s$ is overall successful: the first, second, and third-order confidence intervals are all roughly centered around the true value of $\alpha_s$.
Further, we see that the $\mathcal{O}(\alpha_s^{1})$ confidence interval completely envelopes the $\mathcal{O}(\alpha_s^{2})$ confidence interval, which completely envelopes the $\mathcal{O}(\alpha_s^{3})$ interval.
These results are what we might expect: higher-order RDFs contain more information about the underlying distribution, and thus more closely resemble the target all-orders PDF.
For lower-order calculations, the nuisance parameters have to do less ``work'' to match the target data (since they multiply a lower power of $\alpha_s$), resulting in many more choices of nuisance parameters giving a valid fit.

\begin{figure}[tbh]
    \centering
\includegraphics[width=0.495\linewidth]{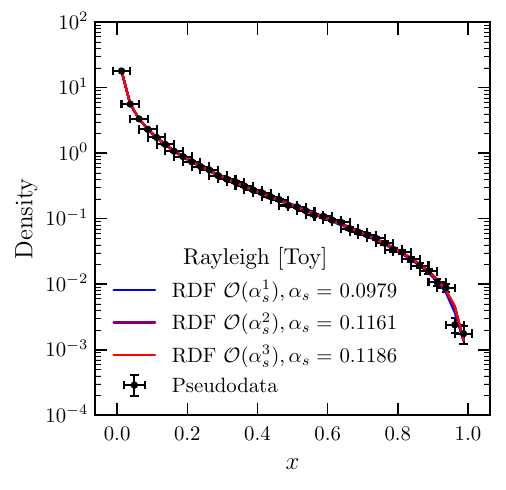}
\includegraphics[width=0.495\linewidth]{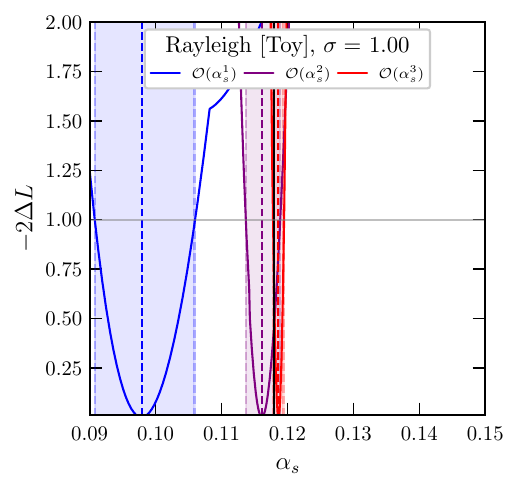}
    \caption{The same as \Fig{exponential_fit}, but with the Rayleigh distribution rather than the exponential distribution.
    }
    \label{fig:rayleigh_fit}
\end{figure}

We show analogous fits to the Rayleigh distribution to first, second, and third order in $\alpha$ in \Fig{rayleigh_fit}.
For this example, while the $\mathcal{O}(\alpha_s^{2})$ and $\mathcal{O}(\alpha_s^{3})$ confidence intervals are roughly centered around the true value of $\alpha_s$, the $\mathcal{O}(\alpha_s^{1})$ confidence interval seems to miss the mark, though it still contains the target within $2\sigma$.
Of course, if well-calibrated, the confidence intervals should only be expected to cover the true value $68\%$ of the time. 
Similar to the exponential toy, lower-order RDFs have more valid choices for the nuisance parameters, resulting in wider confidence intervals.
Further, the $\mathcal{O}(\alpha_s^{2})$ confidence interval does not fully envelope the $\mathcal{O}(\alpha_s^{3})$ interval, although the amount of overlap is high.
We do see that the confidence intervals shrink as the order of the RDF increases.
This shrinkage implies again that embedding more information into the RDF during the matching procedure means that the nuisance parameters can do less work fitting the RDF to the target and instead be more effective at constraining $\alpha_s$.

\subsection{Extracting $\alpha_s$ With Fits to ALEPH Data}\label{sec:thrust_fits}

\begin{figure}[tbh]
    \centering
\includegraphics[width=0.495\linewidth]{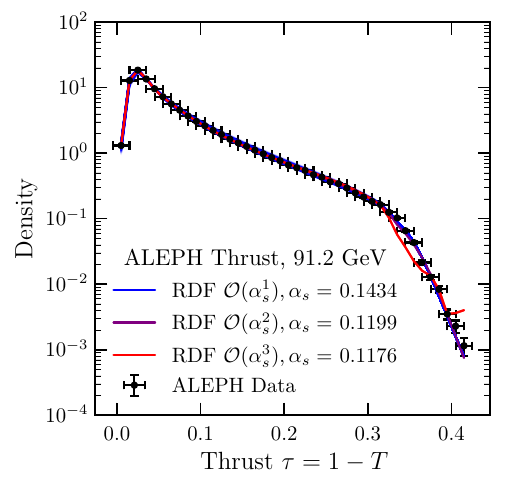}
\includegraphics[width=0.495\linewidth]{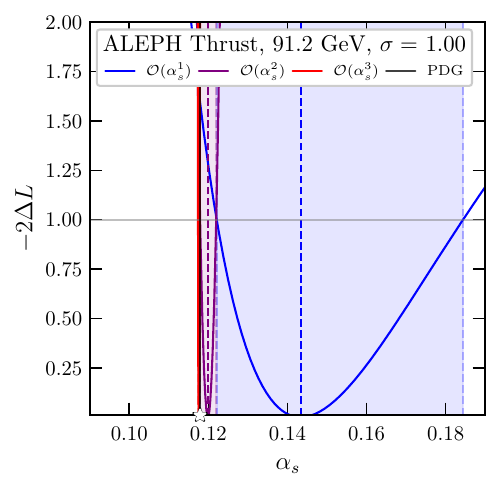}
    \caption{ (Left) RDF fits to 91.2 GeV ALEPH thrust data~\cite{ALEPH:2003obs}. For each of $\O(\alpha_s^1)$ (blue), $\O(\alpha_s^2)$ (purple), and $\O(\alpha_s^3)$ (red), the best-fit RDF is plotted as determined by the fitting procedure for the $\sigma = 1.0$ prior. (Right) The profile-log-likelihood, as defined in \Eq{log_likelihood}, of each of the three orders as a function of $\alpha_s$ for the $\sigma = 1.0$ prior. The minima are indicated by vertical lines, and confidence intervals are drawn where $-2\Delta L = 1$.
    The PDG~\cite{ParticleDataGroup:2024cfk} value of $\alpha_s = 0.1179$ is shown as a black line, with a star on the $x$-axis.
    }
    \label{fig:thrust_fit}
\end{figure}

In this subsection, we use the RDF to perform an extraction of $\alpha_s$ from event shape data.
We use the thrust distribution in $e^+e^-$ collision data collected at $\sqrt{s} = 91.2$ GeV as collected by ALEPH~\cite{ALEPH:2003obs}, which has been made public on HEPData~\cite{hepdata.12794.v1/t54}.
In the left panel of \Fig{thrust_fit}, we show the histogram of measured thrust with statistical uncertainties only.
Following the precedent of other $\alpha_s$ extraction fits~\cite{Kluth:2006bw, Bethke:2006ac,ALEPH:2003obs, OPAL:2011aa, Abbate:2010xh, dEnterria:2022hzv, Benitez:2024nav}, systematic uncertainties are neither shown nor included in our likelihoods.
As this is a proof-of-concept study, for simplicity we will not include the effects of hadronization or experimental systematics in our $\alpha_s$-extraction, though we emphasize that these are essential for a ``real'' extraction.

Similarly to the toy examples, we first use the numeric matching from \Sec{thrust_numeric} to fix the seven learnable arrays up to order $M$.
We then vary over one higher order of the seven learnable arrays as nuisance parameters to extract the value of $\alpha_s$ that maximizes the profile log likelihood ratio.

In \Fig{thrust_fit}, we show the results of extracting $\alpha$ to $\mathcal{O}(\alpha_s^{1})$, $\mathcal{O}(\alpha_s^{2})$, and $\mathcal{O}(\alpha_s^{3})$, with a Gaussian prior on parameters of $\sigma = 1.0$.
We also show the extracted $\alpha_s$ values in \Tab{alpha_thrust} for each order and for $\sigma = 0.5, 1.0$, and $2.0$, as well as the results from \Reference{Becher:2008cf} for comparison.
In the left panel of \Fig{thrust_fit}, we show the RDFs along with the ALEPH data they are fit to, and in the right panel, we show the profile likelihood as a function of $\alpha$ using a Gaussian prior for the nuisance parameters with a standard deviation of 1.0.\footnote{We show the fit results for other choices of the prior standard deviation in \App{other_priors}}

The results for the recovery of $\alpha_s$ are qualitatively similar to those for the Rayleigh toy example.
As expected, the confidence intervals decrease in width as the order in $\alpha_s$ increases.
This implies that if we embed more information into the RDF during the matching procedure, we can make a more precise extraction of $\alpha_s$.
We also plot the PDG world average of $0.1179 \pm  0.0009$~\cite{ParticleDataGroup:2024cfk} for comparison. 
Like in the Rayleigh toy, the $\O(\alpha_s^2)$ and $\O(\alpha_s^3)$ RDF confidence intervals envelope this value, and while the $\O(\alpha_s^1)$ RDF confidence interval doesn't, it gets very close, to just slightly over $1\sigma$.
However, we see that the $\mathcal{O}(\alpha_s^{m})$ interval does not generically envelope the $\mathcal{O}(\alpha_s^{m+1})$ interval, which is undesirable behavior (though their edges touch and they do overlap within 2$\sigma$).
One should keep in mind, however, that our calculations do not account for any type of systematic or hadronization modeling uncertainty.
Given that $1\sigma$ confidence intervals should cover the true (where here, we use the PDG as a proxy) value only $68\%$ of the time, these results are reasonable.
A variant of the right panel of \Fig{thrust_fit} with the vertical axis in log-scale is shown in \Fig{thrust_fit_log} of \App{other_priors}, where one can get a better sense of the shape of the sharp $\O(\alpha_s^3)$ curves.

In \Tab{alpha_thrust}, we compare our $\alpha_s$ extraction results with those from \Reference{Becher:2008cf}, which uses LEP I+II data to fit to a resummed calculation obtained using Soft-Collinear Effective Theory (SCET). 
In addition to testing a Gaussian prior with standard deviation equal to 1.0 for the nuisance parameters, we also extract $\alpha_s$ with two other standard deviations (0.5 and 2.0).
We find that our $\mathcal{O}(\alpha_s^{2})$ and $\mathcal{O}(\alpha_s^{3})$ extractions are consistent with the results from \Reference{Becher:2008cf} when we use a prior with a standard deviation of 1.0, and that the $\O(\alpha_s^1)$ results at least overlap.
Note that \Reference{Becher:2008cf} uses all of LEP runs I and II and uses a variety of energy scales run down to $m_Z$, while we only use ALEPH Run I data solely at $m_Z$, but we do not expect this difference to be the cause of much discrepancy.
We use the entire kinematic range of thrust (possible because the RDF is a proper distribution), while \Reference{Becher:2008cf} uses a limited window, though our results do not change significantly if our fit window is limited. 
In addition, we find that the results for $\alpha_s$ become less prior-dependent as the fit order increases.
These results together imply that if we embed more information into the RDF during the matching procedure, we can make a more accurate and precise extraction of $\alpha_s$.

Given the similar qualitative results to our Rayleigh toy example, it seems likely that the lowest-order RDF does not contain enough information about the target PDF in order for the nuisance parameter minimization to be completely effective.
In the case of $\sigma = 2.0$ for example, the $\O(\alpha_s^1)$ fit fails to find a sensible minimum, and simply returns the largest $\alpha_s$ value we scanned over\footnote{In principle, we could have gone further, but the $\alpha_s$ returned is regardless much larger than one would expect. In general, the $\O(\alpha_s^1)$ fits are finicky, and we had to use an additional L-BFGS minimizer~\cite{Liu1989OnTL} after the gradient descent to avoid getting stuck in local minima.}, and for $\sigma = 0.5$, the $\O(\alpha_s^1)$ result is several standard deviations away from sensible values.
On the other hand, the second-order and higher fits seem relatively robust, and they also do not change much as the prior $\sigma$ changes.
\ul{For users interested in using the RDF framework for parameter fits, we recommend to match at least up to second order before minimizing over nuisance parameters.}

\begin{table}[htb]
  \centering
  \begin{tabular}{lcccc}
    \toprule
    & & \textit{RDF Prior} $\sigma$ & &   \\
    \cmidrule(lr){2-4} 
    \textbf{Order} & $\sigma = 0.5$ & $\sigma = 1.0$ & $\sigma = 2.0$ & LEP I+II (From \Reference{Becher:2008cf}) \\
    \midrule
    Thrust \color{blue}$\O(\alpha_s^1)$         & $0.1555^{+0.0189}_{-0.0139}$ & $0.1434^{+0.0409}_{-0.0212}$ & $^*\textit{0.19}^{\textit{+0.0000}}_{\textit{-0.0595}}$ & $0.1142 \pm 0.0297$  \\
    Thrust \color{violet}$\O(\alpha_s^2)$         & $0.1202^{+0.0010}_{-0.0009}$ & $0.1199^{+0.0019}_{-0.0020}$ & $0.1249^{+0.0067}_{-0.0053}$ & $0.1152 \pm 0.0068$ \\
    Thrust \color{red}$\O(\alpha_s^3)$         & $0.1202^{+0.0002}_{-0.0001}$ & $0.1176^{+0.0002}_{-0.0002}$ & $0.1164^{+0.0003}_{-0.0002}$ & $0.1164 \pm 0.0033$ \\
    \midrule 
    \bottomrule
  \end{tabular}
   \caption{
    Extracted $\alpha_s(m_Z)$ values from the RDF-based fits to ALEPH thrust at 91.2 GeV. The $\sigma = 1.0$ column corresponds to \Fig{thrust_fit}. For comparison, the best fit values and uncertainties for the corresponding orders obtained using SCET fits to ALEPH in Table 4 of \Reference{Becher:2008cf} is shown in the final column. Plots corresponding to the $\sigma = 0.5$ and $\sigma=2.0$ priors are shown in \Fig{thrust_fit_priors} of \App{other_priors}.
    \\
    \textit{*For the $\O(\alpha_s^1)$ thrust fit with $\sigma = 2.0$, the best-fit value occurs at the extreme edge of the considered $\alpha_s$ fit range.}
  }
  \label{tab:alpha_thrust}
\end{table}

\subsection{Brief comparison to other methods}

To conclude this section, we will contrast the RDF nuisance parameter method for parameterizing theory uncertainties with other approaches in the literature.

When calculating physical observables, it is often the case that theoretical uncertainties are dependent on one or more renormalization scales, referred to here heuristically as $\mu_R$.
Physical cross sections may be expressed as 
\begin{equation}
\sigma(Q) = \sum_m^\infty \alpha^m(\mu_R) \sigma_m(Q, \mu_R),
\end{equation}
where $Q$ denotes some physical scale related to the calculation. 
Of course, physical quantities should be independent of the renormalization scale.
However, truncating the sum to a finite order in $\alpha$ will still be $\mu_R$-dependent, so we can use $\mu_R$ as a lever to quantify the uncertainty on the higher-order terms in the sum.
The conventional approach for citing these truncation uncertainties is to quote the physical observable at $\mu_R = Q$ (i.e. the physical scale of the problem), then give an error envelope defined over the variation of the renormalization scale from ${Q/2, 2Q}$.
Despite the widespread use of this conventional approach, it has several drawbacks.
For one, the variation by a factor of 2 is arbitrary.
For another, there is no real probabilistic or physical interpretation of the variation of $\mu_R$ (we cannot, for instance say, that the untruncated cross section has a 95\% chance of lying within the cited envelope), unless this is considered as a subjective Bayesian prior.

As noted by~\Reference{Tackmann:2024kci}, scale variation uncertainties can be insufficient to parameterize the space of all possible higher-order terms --- in a sense, they only capture the physics that we ``already know''.
Our method is \emph{truly} physics-agnostic, as we simply parameterize the form of higher-order coefficients.
Note that one must still choose a finite parameterization for $g$ (e.g. choosing $g$ to be polynomial in $t$), and therefore a bias in the functional form, but in principle a set of complete basis functions may be chosen.
Unlike the approach of \Reference{Tackmann:2024kci}, which also advocates for treating higher-order terms as nuisance parameters, the RDF method \emph{guarantees} that all possible variations are physical for a differential cross section.
An interesting difference to point out between the Theory Nuisance Parameter (TNP) approach in resummed calculations and our method is that while the perturbative objects $f_n$ in the TNP approach (corresponding to our $p_m$ in \Eq{expansion}) take advantage of physics structure (color factors, anomalous dimension structure, etc), the only structure from our RDF approach comes from enforcing unitarity.\footnote{For a similar study extracting $\alpha_s$ from a resummed $Z$ boson $q_T$ spectrum using TNPs, see \Reference{Cridge:2025wwo}.}

The Cacciari-Houdeau approach \cite{Cacciari_2011} estimates truncation uncertainties in a $\mu_R$-independent way.
In particular, given a cross section $\sigma = \sum_{m=0}^\infty c_m \alpha^m$, the authors assume that the coefficients $ c_m$ are independent, but all bounded by a common-but-hidden parameter $\bar{c}$. 
Given a cross section truncated to order $k$, it is possible to estimate higher order coefficients by using Bayes' rule\footnote{Using the notation of Eq (2.22) from \Reference{Bonvini_2020}}

\begin{align}
\begin{split}
P(c_{m+1}|c_1...c_n) = \frac{\int d\bar{c}P(c_{m+1}|\bar{c})P(c_1|\bar{c})...P(c_m|\bar{c})P(\bar{c})}{\int d\bar{c}P(c_1|\bar{c})...P(c_m|\bar{c})P(\bar{c})},
\end{split}
\end{align}
where one must introduce a prior on the hidden parameter $P(\bar{c})$ as well as a dependence of the coefficients on the hidden parameter $P(c_i|\bar{c})$ (taken in the original paper to be flat in the logarithms of the coefficients). 
Thus there is a genuine, prior-motivated notion of uncertainty associated to these coefficients that is not typically found with renormalization-scale estimates.
Later refinements~\cite{Bagnaschi_2015, Bonvini_2020} to the method take into account the expected factorial growth of the coefficients ($ c_m \leq \bar{c} \rightarrow c_m \leq \bar{c}m!$) and  the power growth of $\alpha$ ($\sigma = \sum_{m=0}^\infty c_m \alpha^m \rightarrow  \sum_{m=0}^\infty c_m (\frac{\alpha}{\eta})^m $ for $\eta$ to be determined).

Our approach is similar to Cacciari-Houdeau (and modifications) in that we use prior-dependent bounds on coefficients.
Both our method and the Cacciari-Houdeau method do not rely on the renormalization scale.
However, we make stronger (yet well-motivated) assumptions on the coefficients of the perturbative series  $\sigma = \sum_{m=0}^\infty c_m \alpha^m$ (unitarity) which allow us to write down a closed-form expression for the coefficients --- the RDF itself.
Further, we allow for more user input on the coefficients of the power series: while the Cacciari-Houdeau approach makes claims at the level of $c_m$, our approach allows us to specify the form of $c_m$, for example as a polynomial in $t$.
In the case where there is a strong argument that $c_m$ should take a particular functional form in $t$, our method allows for more fine-grained determination of the associated coefficients.

Other approaches, such as the renormalization-scale dependent model in \Reference{Bonvini_2020} (``Model 2: a new approach using scale variation information"), use renormalization scale information in a more motivated way by promoting the contribution of $\mu_R$ to a learnable parameter, which effectively plays the role of $\bar{c}$ in the Cacciarai-Houdeau method.
Our method does not have any connection to the renormalization scale, although as briefly mentioned in \Sec{resummation}, our framework does not prohibit future modifications that make the RDF dependent on $\mu_R$.

\section{Discussion and Conclusions }
\label{sec:conclusions}

In this paper, we have constructed an ansatz, the \emph{Resummed Distribution Function} (RDF), for parameterizing the set of all higher-order completions of a fixed-order differential cross section consistent with unitarity.
The RDF is capable of matching fixed-order perturbative calculations for the differential cross section of an observable, both analytically and numerically, with higher-order information encoded in the choice of $g$-functions.
With only mild constraints on the $g$-functions (that they are analytic up to single logarithms and bounded-from-above), the differential cross section is guaranteed to be positive, normalized, and finite, as is expected of an all-orders calculation.

We have also demonstrated the utility of the RDF in a number of settings.
In particular, we have shown that the RDF is well-suited for completing QCD observables, either through analytic or numeric matching to a higher-order calculation, and even with multiple observables at once.
We have also used the RDF to simulate a precision measurement by carrying out a mock fit of the strong coupling constant $\alpha_s$ to ALEPH thrust data, showing how the RDF can be use to define a nuisance-parameter based notion of theory uncertainties with explicit priors built in. 
With reasonable prior choices on the nuisance parameters, we were able to extract robust and stable fits for $\alpha_s$ beginning with $\O(\alpha_s^2)$ matched RDFs.

There are a variety of settings where the RDF can be applied.
In cases where it is either difficult or impossible to resum a fixed-order calculation as is the case with simultaneous observables~\cite{Larkoski:2014tva, Lustermans:2019gxu}, such as Energy Flow Polynomials~\cite{Komiske:2017aww, Cal:2022fnm}, the RDF is a method for constraining and reasonably guessing the all-orders structure of the cross section. 
The RDF requires no additional ``physics'' knowledge to be applied (unlike a genuine resummation) other than that the theory predicts valid probabilities.
The built-in method of varying higher-order terms can be used to easily understand theoretical uncertainties, either qualitatively through random variations as was performed in \Secs{toy_examples}{qcd_observables}, or quantitatively with nuisance parameters as was performed in \Sec{nuisance_params}.
In \Sec{nuisance_params}, we showed the utility of the RDF for phenomenology studies by extracting the strong coupling constant from an RDF-matched calculation to thrust.
We might further imagine fitting the RDF to a diverse set of other observables in which resummation effects or theoretical uncertainties can be important, such as $p_T$ tails, electroweak fits, or other event shapes, such as C-parameter~\cite{Hoang:2015hka}.
It would also be interesting to explore bases for $g$ beyond polynomial (or, to potentially replace $g$ with neural networks), which might be more suited to observables with less trivial fixed-order expansions like C-parameter.

Care must be taken when applying and interpreting the RDF. 
As alluded to in \Sec{resummation}, the RDF only approximates all possible \emph{convergent} series, but it is often the case that perturbative calculations in QFT are nonconverging asymptotic series.
Thus, we cannot claim that the ``true'' observable distribution lives within our parameterization, as we are limited only to extrapolating perturbative information.
While the RDF is an all-orders constraint and performs an effective resummation in $\alpha^m p_m$, it is not \emph{the} usual logarithmic resummation one does in QCD.
That is, one should not necessarily expect to get the N$^n$LL resummation of a calculation from the RDF just by applying the matching procedure from \Sec{renormalizing_flow}, though the N$^n$LL resummation is at least contained within some choice of higher-order coefficients. 
We also make no claims whatsoever about nonperturbative physics.
One can consider, for example, augmenting the RDF by convolving with nonperturbative shape functions~\cite{Korchemsky:1999kt} or incorporating nonperturbative information in the numeric matching to potentially alleviate this.
While the RDF can still be applied outside the regime of controlled perturbation theory, in these cases it is only as a regularizer to enforce positivity, normalization, and finiteness, without the nice order-by-order structure in $\alpha$.
Lastly, in defining confidence intervals in \Sec{nuisance_params}, we implicitly assumed Wilks' Theorem.
However, as with any statistical method, when performing a real fit one must calibrate the coverage of these confidence intervals via pseudoexperiments.

Looking to more radical applications of the RDF, we might explore using the multivariable RDF to model the kinematics of an entire event phase space and sample from it in the style of an event generator. 
Alternatively, it would be interesting to explore how to incorporate renormalization scales into the RDF ansatz.
This might be done by making the ansatz contributions $g^*$ and $g_\textrm{Analytic}$ dependent on $\mu_R$, as well as by taking into account the dependence of $\alpha$ on $\mu_R$.
Along the same lines, we could consider factorizing the RDF into hard, soft, and collinear components, with anomalous dimensions satisfying RG equations.
The RDF cannot reproduce the results of resurgence calculations, e.g. essential singularities, but it may be interesting to explore extensions where these are included in some manner.
We hope that other members of the phenomenology community find the RDF to be a useful and complementary tool for their analyses.

\section*{Code and Data}

The code and \texttt{EERAD3v2} data used in this paper are publicly available at \url{https://github.com/rikab/RDF/tree/main}.
All analyses and most plots found in this paper, with the exception of those found in \App{training}, may be reproduced with this repository.
In particular, the code to reproduce the analytic RDF matching studies of \Secs{toy_examples}{qcd_observables} may be found at \url{https://github.com/rikab/RDF/tree/main/analytic}.
The code to reproduce the numeric RDF matching studies of \Secs{toy_examples}{qcd_observables} and the $\alpha_s$ extractions of \Sec{nuisance_params} may be found at \url{https://github.com/rikab/NNEFT/RDF/main/numeric}.
The \texttt{EERAD3v2} thrust data, post-matching parameter fits, and best-fit parameters, are within this same repository.
More details can be found in the \texttt{README} file on the main repository page.
For completeness, we include the full loss and parameter training histories in a separate record available at \url{https://zenodo.org/records/17743471}, which can be used to reproduce the plots \App{training}.

\label{ref:code_data}



\section*{Acknowledgments}

We would like to thank Jesse Thaler for useful discussions and comments, and in particular his offhand comment many years ago, ``A good quantum field theorist is one who knows how to calculate \emph{positive} cross sections'', which partially inspired this work. 
We thank him, Andrew Larkoski, and Benjamin Nachman for feedback on this manuscript.
We would also like to thank Kyle Lee, Jennifer Roloff, and Matt Schwartz for great discussions.

R.G. is supported by The National Science Foundation under grant numbers OAC-2103889, OAC-2411215, and OAC-2417682, and by the U.S. DOE Office of High Energy Physics under grant number DE-SC1019775.
A significant portion of this work was completed under NSF Cooperative Agreement PHY-2019786 (The NSF AI Institute for Artificial Intelligence and Fundamental Interactions, \url{http://iaifi.org/}), and U.S. DOE Office of High Energy Physics grant number DE-SC0012567. 
R.M. received support through Schmidt Sciences, LLC.
This research used resources of the National Energy Research Scientific Computing Center, a DOE Office of Science User Facility supported by the Office of Science of the U.S. Department of Energy under Contract No. DE-AC02-05CH11231 using NERSC award HEP-ERCAP0021099.

\appendix

\section{Supplementary plots}
\label{app:supp_plots}

In this appendix, we display supplementary plots that augment the main text.
In \App{other_priors}, we show the $\sigma = 0.5$ and $\sigma = 2.0$ variants of the thrust fit likelihoods from \Sec{thrust_fits}.
We also discuss using flat priors.
In \App{training}, we show the minimization history corresponding to the $\O(\alpha_s^1)$ exponential toy in \Sec{numeric_matching_toy}.

\subsection{Supplementary Plots for Thrust Fits}\label{app:other_priors}

\begin{figure}[tbh]
    \centering
\includegraphics[width=0.495\linewidth]{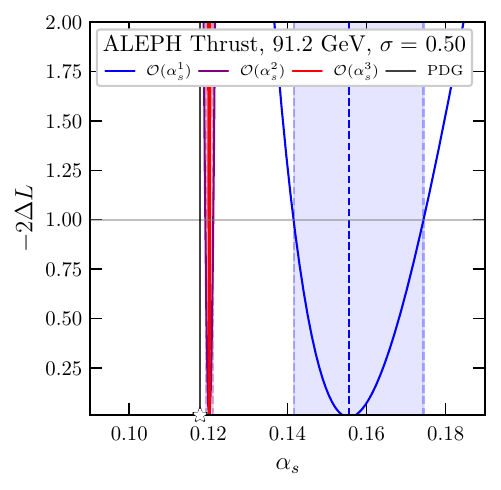}
\includegraphics[width=0.495\linewidth]{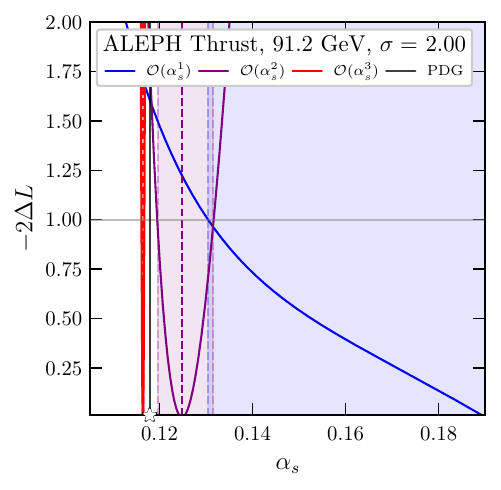}
    \caption{The same as the right panel of \Fig{thrust_fit}, but with a Gaussian prior scale on $g_{mn}$ of $\sigma = 0.5$ (left) and $\sigma = 2.0$ (right) instead of $\sigma = 1.0$.}
    \label{fig:thrust_fit_priors}
\end{figure}

In \Fig{thrust_fit_priors}, we show the analogue of \Fig{thrust_fit}, but with $\sigma = 0.5$ and $\sigma = 2.0$ as shown in \Tab{alpha_thrust}.
While the $\mathcal{O}(\alpha_s^2)$ and $\mathcal{O}(\alpha_s^3)$ results are roughly consistent between all the prior choices, the $\mathcal{O}(\alpha_s^1)$ calculation fit is not.
They are much greater than the expected value of $\alpha_s \approx 0.118$, and in fact, the $\sigma = 2.0$ fit reaches the edge of our fit range.

\begin{figure}[tbh]
    \centering
\includegraphics[width=0.495\linewidth]{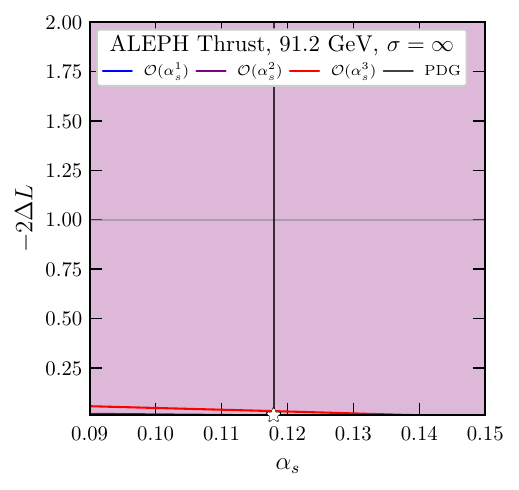}
    \caption{The same as the right panel of \Fig{thrust_fit}, but with a completely flat prior.}
    \label{fig:thrust_fit_inf}
\end{figure}

In \Fig{thrust_fit_inf}, we show the result of having \emph{no} prior term (equivalently, $\sigma = \infty$) in \Eq{log_likelihood}.
As expected, the confidence intervals are ``infinitely'' wide, at least within the reasonable $\alpha_s$ range considered.
The only reason why we achieve nonzero values of the likelihood at all is that there is still a very mild prior that comes from our nuisance parameterization being finite.
These plots together make it clear that priors are a desirable feature in defining theory uncertainties.

\begin{figure}[tbh]
    \centering
\includegraphics[width=0.495\linewidth]{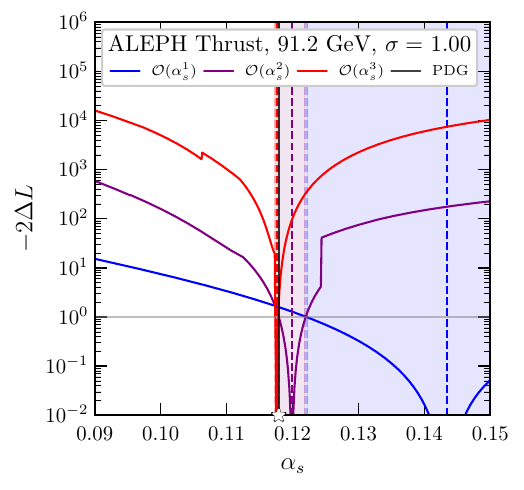}
\includegraphics[width=0.495\linewidth]{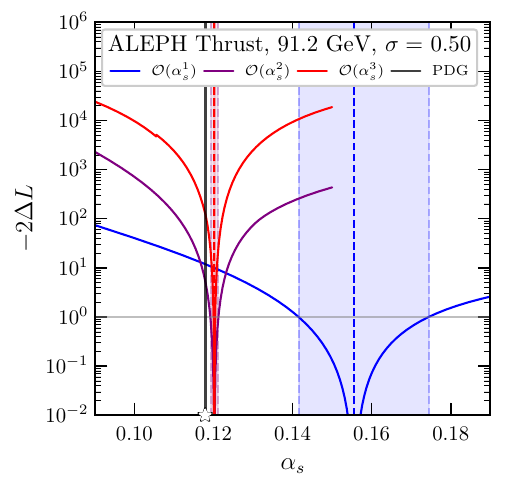}
\includegraphics[width=0.495\linewidth]{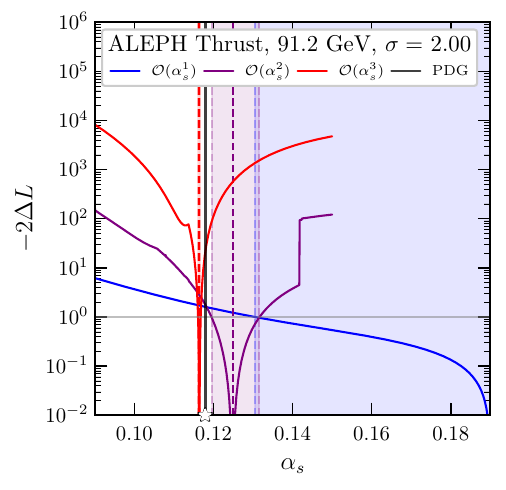}
\includegraphics[width=0.495\linewidth]{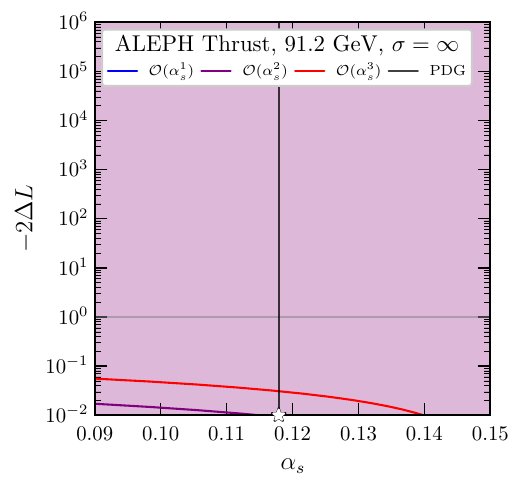}
    \caption{The same as \Fig{thrust_fit}, \Fig{thrust_fit_priors}, and \Fig{thrust_fit_inf}, but with the vertical axis in log scale.}
\label{fig:thrust_fit_log}
\end{figure}

Finally in \Fig{thrust_fit_log}, for visual clarity, we remake all likelihood plots (\Fig{thrust_fit}, \Fig{thrust_fit_priors}, and \Fig{thrust_fit_inf}) with the vertical axis in log scale. 
This is to better show the full magnitude of the likelihood and give a better sense of the curvature, especially at $\O(\alpha_s^3)$.

\subsection{Training Plots}\label{app:training}

In this appendix, we show an example of the numeric matching procedure.
The purpose of this appendix is primarily to give a semi-pedagogical taste of the numeric matching for readers unfamiliar with this machine-learning style minimization by attaching some real numbers to each of the parameter objects.
We elect only to show the $\O(\alpha_s^1)$ exponential matching, corresponding to the top row of \Fig{exponential_matching}. 
This is for several reasons: first and foremost, the actual values of the $g$ matrices are not necessarily informative, except in the toys where we can check if they match the analytic expectations.
One way to see this is that the $T$-parameters can be slightly degenerate with the $g$ parameters, making both more complex to interpret.
Second, the $g$ matrices can be rather large beyond $\O(\alpha_s^1)$, and there is little information to gain by staring at a large number of minimization curves.

\begin{figure}
    \centering
\includegraphics[width=0.435\linewidth]{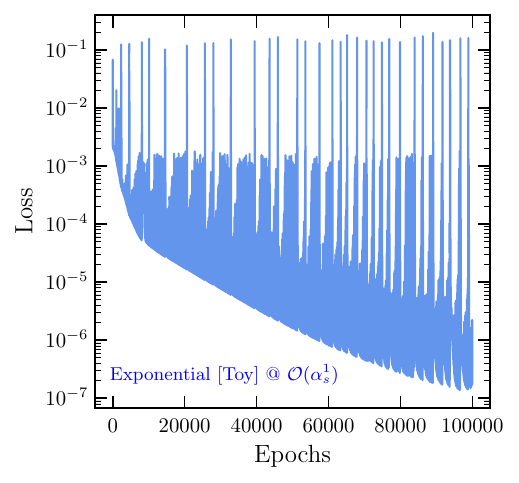}
\includegraphics[width=0.435\linewidth]{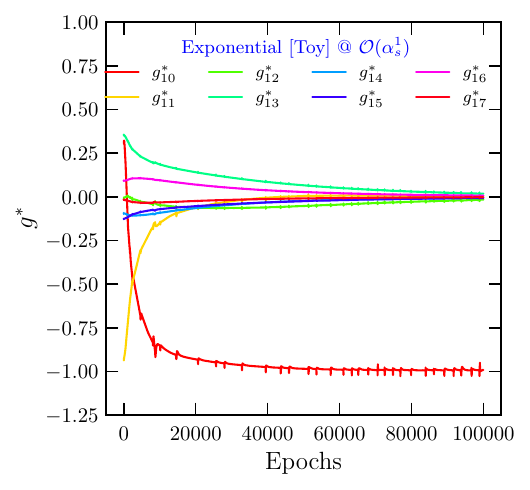}
\includegraphics[width=0.435\linewidth]{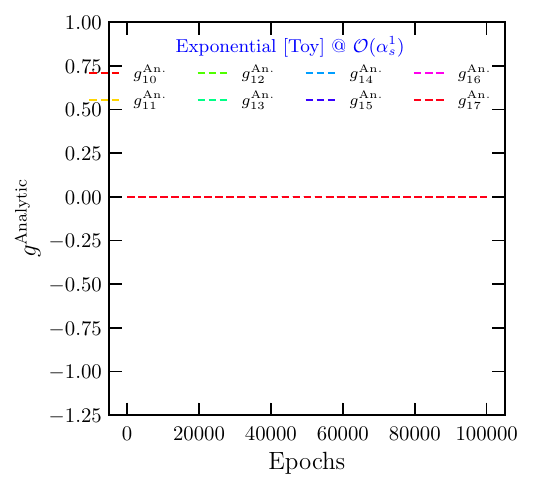}
\includegraphics[width=0.435\linewidth]{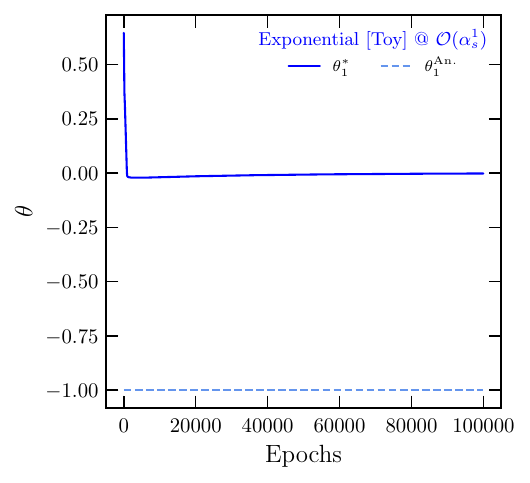}
\includegraphics[width=0.435\linewidth]{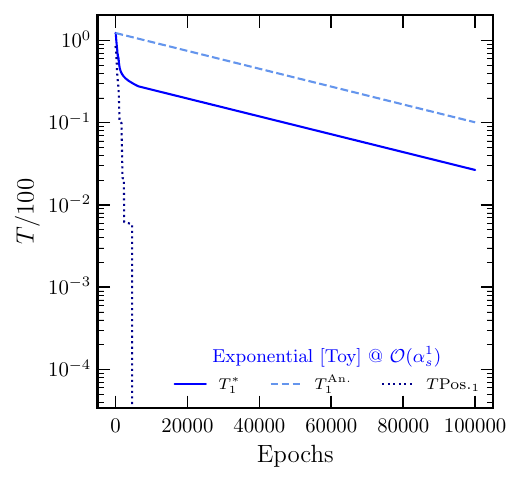}
    \caption{Training history for the $\O(\alpha_s^1)$ matching to the exponential toy, corresponding to the top row of \Fig{exponential_matching}. The plots correspond to: (top left) the overall loss, as defined in  \Eq{numeric_mse_loss}, (top right) the $g^*$ matrix, (middle left) the $g_{\rm Analytic}$ matrix, (middle right) $\theta^*$ and $\theta_{\rm Analytic}$, and (bottom) $T^*$, $T_{\rm analytic}$, and $T_{abs}$ (scaled by a factor of 100).
    }
    \label{fig:training}
\end{figure}

The training history is plotted in \Fig{training}.
As is expected, the loss converges to zero.
All elements of the $g^*$ matrix converge to zero, as well as the $\theta$ and $T$ parameters, except for $g_{10} \to -1.0$, which is precisely the expectation of \Eq{exponential_g}.
Note that while the loss continues to trend downwards, the analytic coefficients do not appreciably change over the course of training (except for changes due to weight decay in the case of $T_{\rm analytic}$), as they do not contribute at $O(\alpha_s^1)$.

\section{Numeric Matching Details}\label{app:numerics}

In this appendix, we provide additional details about our initialization procedure and minimization procedure for fits.
In \App{initialization}, we discuss the procedure to initialize our (nuisance) parameters.
In \App{min}, we provide more details about the minimization procedure used for the fits in \Sec{nuisance_params}.
Finally, in \App{training_hyperparams}, we show the training hyperparameters for the numeric matching procedure.

\subsection{Reroll initialization procedure}\label{app:initialization}

To initialize the seven learnable objects $\phi$ (and later, the nuisance parameters $\nu$), we define a \textit{reroll procedure}, with the purpose of getting a good ``head start'' for gradient descent. 
This procedure is analogous to simulated annealing.
For a fit to order $m$: we first freeze all the coefficients corresponding to the $m-1$ fit to the values recovered during that lower order matching procedure (if available).
Then, for a fixed number of epochs $N$ (we choose 1000), we randomly resample all the non-fixed parameters from Gaussian distributions centered on the original values. 
The standard deviation $\sigma$ is given by 
\begin{align}
\sigma = 0.1 \times \frac{N - \rm{epoch}_i}{N} \times \frac{1}{\sqrt{\rm{counts} + 1}},
\end{align}
which decreases linearly to zero over the entire initialization procedure, and further decreases each time a new best loss is found (the number of times is given by ``counts"). 
We run 1024 of these initialization threads in parallel, and we take the set of parameters with the lowest loss at the end as our initialization for gradient descent.

These choices are mostly ad-hoc, and were determined primarily by trial-and-error and should not be misconstrued as an optimal choice. \
We find that this procedure is empirically useful for finding good minima.

\subsection{Minimization Details}\label{app:min}

In order to obtain our profile likelihood, we must minimize \Eq{profile} over $\nu$ for many values of $\alpha_s$.
We scan over 600 values of $\alpha_s$ evenly between [0.09, 0.15] (although for some of the $\O(\alpha_s^1)$ fits, this range is insufficient and we extend to 1000 values between [0.09, 0.19]).
Performing 600-1000 individual fits is expensive, so we take advantage of the fact that the likelihood is expected to be continuous in $\alpha_s$: we initialize the fit for each $\alpha_s$ using the final fit parameters of the previous $\alpha_s$, with the assumption that the minimum is likely closer.
We scan from left-to-right in $\alpha_s$, then back again from right-to-left to avoid any potential hysteresis effects.
We take the minimum of the two runs: usually, the left-to-right run has large downwards jumps as there are still local minima to jump out of.

For each $\alpha_s$ after the initial fit (described below), we perform the following 3-step minimization procedure, which we have empirically found gives us good minima:
\begin{enumerate}
    \item \textbf{Reroll}: We initialize the fit to the best parameters of the previous $\alpha_s$. We then run the reroll procedure of \App{initialization}, but for speed only use 250 epochs and only 64 parallel runs. This is to ``jostle'' the parameters away from the previous $\alpha_s$.
    \item \textbf{Train}: We perform gradient descent for $2.5 \times 10^4$ epochs, similar to the matching procedure.
    \item \textbf{Refine}: We use the ``limited-memory'' L-BFGS~\cite{Liu1989OnTL} minimizer, which is an approximate second-order-minimizer to refine the results. This can occasionally cause jumps out of local minima, which is why the second right-to-left run is important. 
\end{enumerate}
For the initial value of $\alpha_s = 0.09$, we do the same 3 steps, except we run the reroll initializer for 1000 epochs with 1024 parallel runs, and the training runs for 2M epochs.

As a final note, we find that float precision is empirically important.
When using \texttt{float64} precision rather than \texttt{float32}, we find that the refining step is much more likely to jump out of a global minimum.
Thus, we recommend using \texttt{float64} whenever possible.

\subsection{Training hyperparameters}
\label{app:training_hyperparams}

In \Tab{hyperparams}, we show the training hyperparameters used for all of the numeric fits shown in this work.
The hyperparameters are generally the same for all training runs, though we sometimes make minor adjustments to the number of training epochs and learning rate to ensure convergence of the $g$ coefficients.
For all plots, we take the coefficients at the epoch of lowest loss.

\begin{table}[htb]
\centering
\begin{tabular}{llrrrrr}
\toprule
                             &    & Maximum $t$ power & Epochs & Learning Rate & Weight Decay &  \\
                             \midrule
\multirow{3}{*}{Exponential} & $\mathcal{O}(\alpha^1_s)$ & 7           & 2$\times10^5$    & 5$\times10^{-4}$          & 1$\times10^{-3}$         &  \\
                             & $\mathcal{O}(\alpha^2_s)$ & 7           & 2$\times10^5$    & 5$\times10^{-4}$          & 1$\times10^{-3}$         &  \\
                             & $\mathcal{O}(\alpha^3_s)$ & 7           &      1$\times10^{6}$   &   2$\times10^{-4}$            & 1$\times10^{-3}$         &  \\
                             \midrule
\multirow{3}{*}{Rayleigh}    & $\mathcal{O}(\alpha^1_s)$ & 7           & 1$\times10^5$    & 5$\times10^{-4}$          & 1$\times10^{-3}$         &  \\
                             & $\mathcal{O}(\alpha^2_s)$ & 7           & 1$\times10^5$    & 5$\times10^{-4}$          & 1$\times10^{-3}$         &  \\
                             & $\mathcal{O}(\alpha^3_s)$ & 7           &      1$\times10^{6}$   &    5$\times10^{-4}$           & 1$\times10^{-3}$         &  \\
                             \midrule
\multirow{3}{*}{Thrust}      & $\mathcal{O}(\alpha^1_s)$ & 7           & 1$\times10^5$    & 2$\times10^{-3}$          & 1$\times10^{-3}$         &  \\
                             & $\mathcal{O}(\alpha^2_s)$ & 7           & 1$\times10^6$    & 2$\times10^{-3}$          & 1$\times10^{-3}$         &  \\
                             & $\mathcal{O}(\alpha^3_s)$ & 7           & 1$\times10^6$    & 2$\times10^{-3}$          & 1$\times10^{-3}$         &  \\
\midrule
\bottomrule
 \end{tabular}
    \caption{
   Training hyperparameters for all numeric matching examples. The matching procedure is defined in \Sec{numeric_matching}.
  }
  \label{tab:hyperparams}
\end{table}

\bibliographystyle{JHEP}
\bibliography{refs, jet_refs}

\end{document}